
\input harvmac.tex


\def\unlockat{\catcode`\@=11}

\def\lockat{\catcode`\@=12}

\unlockat


\def\newsec#1{\global\advance\secno by1\message{(\the\secno. #1)}
\global\subsecno=0\global\subsubsecno=0
\global\deno=0\global\prono=0\global\teno=0\eqnres@t\noindent
{\bf\the\secno. #1}
\writetoca{{\secsym} {#1}}\par\nobreak\medskip\nobreak}
\global\newcount\subsecno \global\subsecno=0
\def\subsec#1{\global\advance\subsecno
by1\message{(\secsym\the\subsecno. #1)}
\ifnum\lastpenalty>9000\else\bigbreak\fi\global\subsubsecno=0
\global\deno=0\global\prono=0\global\teno=0
\noindent{\it\secsym\the\subsecno. #1}
\writetoca{\string\quad {\secsym\the\subsecno.} {#1}}
\par\nobreak\medskip\nobreak}
\global\newcount\subsubsecno \global\subsubsecno=0
\def\subsubsec#1{\global\advance\subsubsecno by1
\message{(\secsym\the\subsecno.\the\subsubsecno. #1)}
\ifnum\lastpenalty>9000\else\bigbreak\fi
\noindent\quad{\secsym\the\subsecno.\the\subsubsecno.}{#1}
\writetoca{\string\qquad{\secsym\the\subsecno.\the\subsubsecno.}{#1}}
\par\nobreak\medskip\nobreak}

\global\newcount\deno \global\deno=0
\def\de#1{\global\advance\deno by1
\message{(\bf Definition\quad\secsym\the\subsecno.\the\deno #1)}
\ifnum\lastpenalty>9000\else\bigbreak\fi
\noindent{\bf Definition\quad\secsym\the\subsecno.\the\deno}{#1}
\writetoca{\string\qquad{\secsym\the\subsecno.\the\deno}{#1}}}

\global\newcount\prono \global\prono=0
\def\pro#1{\global\advance\prono by1
\message{(\bf Proposition\quad\secsym\the\subsecno.\the\prono #1)}
\ifnum\lastpenalty>9000\else\bigbreak\fi
\noindent{\bf Proposition\quad\secsym\the\subsecno.\the\prono}{#1}
\writetoca{\string\qquad{\secsym\the\subsecno.\the\prono}{#1}}}

\global\newcount\teno \global\prono=0
\def\te#1{\global\advance\teno by1
\message{(\bf Theorem\quad\secsym\the\subsecno.\the\teno #1)}
\ifnum\lastpenalty>9000\else\bigbreak\fi
\noindent{\bf Theorem\quad\secsym\the\subsecno.\the\teno}{#1}
\writetoca{\string\qquad{\secsym\the\subsecno.\the\teno}{#1}}}
\def\subsubseclab#1{\DefWarn#1\xdef
#1{\noexpand\hyperref{}{subsubsection}%
{\secsym\the\subsecno.\the\subsubsecno}%
{\secsym\the\subsecno.\the\subsubsecno}}%
\writedef{#1\leftbracket#1}\wrlabeL{#1=#1}}

\lockat

\def\IB{\relax\hbox{$\inbar\kern-.3em{\rm B}$}}
\def\IC{\relax\hbox{$\inbar\kern-.3em{\rm C}$}}
\def\ID{\relax\hbox{$\inbar\kern-.3em{\rm D}$}}
\def\IE{\relax\hbox{$\inbar\kern-.3em{\rm E}$}}
\def\IF{\relax\hbox{$\inbar\kern-.3em{\rm F}$}}
\def\IG{\relax\hbox{$\inbar\kern-.3em{\rm G}$}}
\def\IGa{\relax\hbox{${\rm I}\kern-.18em\Gamma$}}
\def\IH{\relax{\rm I\kern-.18em H}}
\def\IK{\relax{\rm I\kern-.18em K}}
\def\IL{\relax{\rm I\kern-.18em L}}
\def\IP{\relax{\rm I\kern-.18em P}}
\def\IR{\relax{\rm I\kern-.18em R}}
\def\IZ{\relax\ifmmode\mathchoice
{\hbox{\cmss Z\kern-.4em Z}}{\hbox{\cmss Z\kern-.4em Z}}
{\lower.9pt\hbox{\cmsss Z\kern-.4em Z}}
{\lower1.2pt\hbox{\cmsss Z\kern-.4em Z}}\else{\cmss Z\kern-.4em Z}\fi}

\def\II{\relax{\rm I\kern-.18em I}}

\def\frac#1#2{{#1\over#2}}



\def\jb{{\bar j}}


\def\imp{$\Rightarrow$}

\def\inbar{\,\vrule height1.5ex width.4pt depth0pt}
\font\cmss=cmss10 \font\cmsss=cmss10 at 7pt


\font\manual=manfnt \def\dbend{\lower3.5pt\hbox{\manual\char127}}


\def\boxit#1{\vbox{\hrule\hbox{\vrule\kern8pt
\vbox{\hbox{\kern8pt}\hbox{\vbox{#1}}\hbox{\kern8pt}}
\kern8pt\vrule}\hrule}}
\def\mathboxit#1{\vbox{\hrule\hbox{\vrule\kern8pt\vbox{\kern8pt
\hbox{$\displaystyle #1$}\kern8pt}\kern8pt\vrule}\hrule}}

\Title{ \vbox{\baselineskip12pt\hbox{hep-th/9905032}
\hbox{YCTP-P7-99 }
\hbox{SPIN-1999/xx}
\hbox{INLO-PUB-10/99}
}}
{\vbox{
\centerline{Two-dimensional Conformal Field Theories on}
\bigskip
\centerline{$AdS_{2d+1}$ Backgrounds}
 \centerline{}}}
\medskip
\centerline{
Jan de Boer $^{1,2}$, and Samson L. Shatashvili $^{3}$\footnote{*}{On
leave of
absence from St. Petersburg Branch of Steklov Mathematical Institute,
 Fontanka,
St.
Petersburg,
Russia.}}

\vskip 0.5cm
\centerline{$^{1}$ Spinoza Institute, University of Utrecht,}
\centerline{ Leuvenlaan 4, 3584 CE Utrecht, The Netherlands}
\centerline{}
\centerline{$^{2}$ Instituut-Lorentz for Theoretical Physics,
University of Leiden,}
\centerline{PO Box 9506, NL-2300 RA, Leiden, The Netherlands}
\centerline{}
\centerline{$^3$ Yale University, Department of Physics, P. O. Box 208120,
New
Haven, CT 06520-8120 }
\centerline{}
\vskip 0.1cm
Various exact two-dimensional
conformal field theories with $AdS_{2d+1}$ target space
are constructed. These models
can be solved using bosonization techniques.
Some examples are presented that can be
used in building perturbative
superstring theories with $AdS$ backgrounds, including
$AdS_5$.

\medskip
\noindent

\Date{May 5, 1999}

\newsec{Introduction}

The dualities \ref\mal{J. Maldacena, ``The Large $N$ Limit of Superconformal
 Field Theories and Supergravity,'' Adv. Theor. Math. Phys. {\bf 2} (1998) 231,
hep-th/9711200.} 
\ref\gkp{S. Gubser, I. Klebanov and A. Polyakov, ``Gauge Theory Correlators from
Non-Critical String Theory,'' hep-th/9802109.}
\ref\w{E. Witten, ``Anti De Sitter Space and Holography,'' hep-th/9802150.}
between certain conformal field theories and string
theories on anti-de Sitter (AdS) spaces with RR fluxes have led
to renewed interest in the description of 2d conformal field theories
with RR backgrounds. A useful description of the latter would allow one
to go beyond the supergravity approximation and study $1/N$
and $1/g^2 N$ corrections in the field theory.

The description of RR backgrounds in the NSR formulation is a difficult
problem. It is not even known how to describe a condensate of RR vertex
operators at the classical level (for a recent attempt, see
\ref\bele{D. Berenstein and R.G. Leigh, ``Superstring Perturbation
Theory and Ramond-Ramond Backgrounds,'' hep-th/9904104.}).
On the other hand, the classical description of $AdS_5 \times S^5$
and $AdS_3 \times S^3$ backgrounds with RR fields in the Green-Schwarz
formalism is known
\ref\several{R.R. Metsaev and A.A. Tseytlin, ``Type IIB Superstring Action
In $AdS_5 \times S^5$ Background,'' Nucl. Phys. {\bf B533} (1998) 109,
hep-th/9805028;
R. Kallosh, J. Rahmfeld and A. Rajaraman, ``Near Horizon Superspace,''
JHEP {\bf 09} (1998) 002, hep/th-9805217;
I. Pesando, ``A $\kappa$ Gauge Fixed Type IIB Superstring Action of
$AdS_5 \times S^5$,'' JHEP 11 (1998) 002, hep-th/9808020;
R. Kallosh and J. Rahmfeld, ``The GS String Action of $AdS_5 \times S^5$,''
Phys. Lett. {\bf B443} (1998) 143, hep-th/9808038;
I. Pesando, ``The GS Type IIB Superstring on $AdS_3 \times S^3 \times T^4$,''
hep-th/9809145,
J. Rahmfeld and A. Rajaraman, ``The GS String Action on $AdS_3 \times S^3$,''
hep-th/9809164;
J. Park and S.-J. Rey, ``Green-Schwarz Superstring $AdS_3 \times S^3$,''
hep-th/9812062.}.
The quantization of the GS string is more complicated than the quantization
of the NSR string. In flat space, the GS string can be quantized in light-cone
gauge, which breaks manifest target space covariance. This is no longer
possible on curved space like $AdS$; quantization in other gauges has
been discussed in \ref\quant{M. Yu and B. Zhang, ``Light-Cone Gauge Quantization
of String Theories on $AdS_3$,'' hep-th/9812216;
A. Rajaraman and M. Rozali, ``On the Quantization of the GS String on $AdS_5
\times S^5$,'' hep-th/9902046.}

A third, more promising approach is to employ GS type variables for the
NSR string, which combines the covariance of the GS approach with
the quantizability of the NSR string. Such variables
exist in four \ref\berfour{N. Berkovits, ``Covariant
Quantization of the Green-Schwarz Superstring in a
Calabi-Yau Background,'' Nucl. Phys. {\bf B431} (1994) 258, hep-th/9404162.}
and six \ref\berksix{N. berkovits and C. Vafa,
``$N=4$ Topological Strings,'' Nucl. Phys. {\bf B433} (1995) 123, hep-th/9407190.}
dimensions. In ten dimensions variables that
preserve $U(5)\subset SO(10)$ super Poincar\'e invariance
exist \ref\berten{N. Berkovits, ``Quantization of the
Superstring with Manifest $U(5)$ Super-Poincar\'e
Invariance,'' hep-th/9920099.}. The six-dimensional
variables can be used to find a description of superstrings
on $AdS_3 \times S^3$ with both RR and NS background
fields turned on \ref\bvw{N. Berkovits, C. Vafa and E. Witten,
``Conformal Field Theory of $AdS$ Background with Ramond-Ramond
 Flux,'' hep-th/9902098.}

The case of $AdS_3 \times S^3$ with RR background is special because
the theory can be S-dualized to $AdS_3 \times S^3$ with only
NS fields turned on, which in turn can be described in a
straightforward way using WZW models. The resulting theory
and the relation between the space-time and world-sheet CFTs
were studied in \ref\gks{A. Giveon, D. Kutasov and N. Seiberg,
``Comments on String Theory on $AdS_3$,'' Adv. Theor. Math. Phys.
{\bf 2} (1998) 733, hep-th/9806194.}, \ref\bort{
J. de Boer, H. Ooguri, H. Robins and J. Tannenhauser,
``String Theory on $AdS_3$,'' JHEP {\bf 12} (1998) 026,
hep-th/9812046.}, \ref\ks{
D. Kutasov and N. Seiberg, ``More Comments on String Theory on $AdS_3$,''
hep-th/9903219.}. We believe that the $AdS_5$ and $AdS_7$ CFTs
described in this paper can be studied in a similar fashion.

In general, to construct a world-sheet description of a given
string background one needs to find a 2d CFT that describes that
background, check it has the correct central charge, and couple
it to ghosts (or ghost-like variables depending on the
approach). Finding exact conformal sigma models is not an
easy problem. The models that have been most extensively
studied and that can be solved exactly are rational conformal
field theories. Many of those correspond to WZW theories (or
their cosets). The corresponding Lagrangians are two-dimensional
sigma models on group manifolds that include WZ terms. A priori
the coefficients in front of the kinetic and the WZ term are
unrelated, but for most groups conformal invariance requires
them to be proportional to each other. An exception are the
supergroups $PSL(n|n)$, for which it has been shown recently
that the theory is conformal for arbitrary values of the two
parameters \bvw, \ref\bzv{M. Bershadsky, S. Zhukov and
A. Vaintrob, ``$PSL(n|n)$ Sigma Model as a Conformal Field Theory,''
hep-th/9902180.}. In fact, the description in \bvw\ of
superstrings on $AdS_3 \times S^3$ is based on a sigma model
with target space the group manifold $PSL(2|2)$, coupled
in a certain way to ghost fields.
Together with the physical state condition this provides a complete
world-sheet description of $AdS_3 \times S^3$
background with an arbitrary amount of $RR$ and $NS$ fluxes.

In this paper we construct exact CFTs with $AdS_{2d+1}$ backgrounds
and compute their central charges. As far as we know no previous
examples of exact quantum CFTs with $AdS_{2d+1}$ target space
($d>1$) were known, independently of whether RR background
fields are turned on or not. The exact CFTs we construct all
correspond to a novel type of cosets of WZW models, and
depend on one free parameter. Standard
cosets cannot be used, because their target spaces are
generically singular spaces of the form $G/H$ where $H$ acts
on $G$ as $g\rightarrow h^{-1} g h$. Homogeneous spaces
like $AdS_d$, on the other hand,
are of the form $G/H$ where $H$ acts from the left
or right on $G$, and gauging such a subgroup in WZW
models is anomalous.

Let us list some results:

\vskip 0.1cm

1. For the purely bosonic case we claim that the sigma model

\eqn\firstclaim{S_{AdS_{2d+1}}={k \over 2\pi}
\int\,d^2 z\,(\partial\phi\bar{\partial}\phi+e^{2\phi}\partial
\bar{\gamma}_r \bar{\partial} \gamma_r)
+{1 \over 2 \pi} \int \, d^2 z \, (d-1) \phi \sqrt{g} R
, \quad r=1\ldots d}
is exactly conformal. It has $AdS_{2d+1}$ as target space, a non-zero
$B$-field and dilaton and Virasoro central charge:
\eqn\cforone{
c= (2d+1) + {6 \over k-2 d} . }
The non-zero $B$-field  depends on a choice of complex structure
on the boundary of $AdS_{2d+1}$ and therefore breaks the Lorentz group
from $SO(2d)$ to $U(d)$. It is an interesting question whether
a supersymmetric version of this theory 
(see below)
can be mapped (using a combination of $T$- and $S$-dualities)
to $AdS_5$ with only the RR five-form turned on.

2. One can also consider the novel type of cosets for supergroups. The
resulting sigma models will contain
anti-commuting (world-sheet scalar) variables $\theta$.
 For example, for the case of $SL(N|N)/SL(N-1)^2$ the resulting sigma
model contains $2N^2$ real anti-commuting fields. The bosonic part
of the sigma model has as metric $AdS_{2N-1} \times AdS'_{2N-1}$
with a nonzero $B$-field and dilaton. The prime on $AdS'$ indicates
that the metric is that of $AdS$ but with a wrong sign for $d\phi^2$,
$ds'^2 = -d\phi^2  + e^{2\phi} |d\gamma_r|^2$. Besides a kinetic term
the anti-commuting variables also have a fermionic $B$-field. The
precise meaning of this fermionic $B$-field depends on the way the
sigma model is promoted to a full-fledged string theory (i.e. on
the coupling to the ghost degrees of freedom). Although this will
not be discussed in this paper, we suspect that there is a close
relation between the fermionic $B$-field and RR background fields.
The central charge of the $SL(N|N)/SL(N-1)^2$ sigma model is
$c=-2N^2 +4N-2$. For $N=3$ we get a ten-dimensional space-time
with $c=-8$.

Another interesting example is $SL(4|4)/SP(2)^2$. The corresponding
homogeneous superspace plays an important role in the construction
of the action for the GS string on $AdS_5 \times S^5$. In our case
we find as space-time metric $AdS_5 \times AdS_5'$ 
(we will comment on the relation between
$AdS_5'$ and $S^5$ later), and 32 anti-commuting
world-sheet scalars. Furthermore,
the central charge of this theory is $c=-22$. It is tempting
to conjecture a relation to the GS type of approach; however, our sigma
model does not appear to have an analogue of kappa symmetry\foot{We 
thank M. Bershadsky for discussions on the bosonization
of current algebras for supergroups and related questions.}.

3. If we combine the examples given under 1 with their
 analytic continuations
from $\phi$ to $i\phi$
(which corresponds to changing the geometry of $AdS$ to that of
$AdS'$)
we get bosonic conformal sigma models with target space
$AdS_{2d_1+1} \times AdS'_{2d_2+1}$. If $d_1+d_2=4$ and
the two levels $k_{1,2}$ of the two individual sigma models
satisfy $k_1-2d_1=-(k_2-2 d_2)$ the central charge is exactly
equal to ten. By adding NSR type world-sheet fermions we
can construct a theory with $N=1$ world-sheet superconformal
invariance and central charge $c=15$. The number of unbroken
target space supersymmetries equals $(d_1+1)(d_2+1)$.

We will bosonize all models discussed above, thereby
showing that they are ``exactly solvable''. We put the
exact solvability between quotation marks because we expect
to encounter the same difficulties that one meets when one tries
to solve the noncompact $SL(2,R)$ WZW model using bosonization.

\vskip 0.02cm

The outline of this paper is as follows. In section~2 we describe the
bosonic sigma models with target space $AdS_{2d+1}$ and show
that the corresponding background fields solve
(the lowest order in $\alpha'$) supergravity
equations of motion.
We will also discuss the
global and local symmetries of the ``cosets'' and comment on
their holographic nature.

In section~3 we
describe how such models can be obtained by means of a novel kind
of cosets of WZW models. This procedure was first hinted in
\ref\ff{
B.L.~Feigin, E.V.~Frenkel,
The family of representations of affine Lie algebras,
Usp.\ Mat.\ Nauk.\ {\bf 43}, 227-228 (1988),
Russ.\ Math.\ Surv.\ {\bf 43}, 221--222 (1989);

Affine Kac-Moody algebras and semi-infinite flag manifolds,
Commun.\ Math.\ Phys.\ {\bf 128}, 161--189 (1990);

Representations of affine Kac-Moody algebras, bosonization and
resolutions,
Lett.\ Math.\ Phys.\ {\bf 19},  307-317 (1990);

Representations of affine Kac-Moody algebras and bosonization,
pp.\ 271--316 in: Physics and Mathematics of Strings,
eds.~L.\ Brink at al,
World Scientific, Singapore, 1990;

E.~Frenkel,
 Free field realizations in representation theory and conformal
field theory,
in: Proceedings of the ICM, Z\"urich 1994,
preprint: hep-th/9408109.}
\ref\gmmos{A. Gerasimov, A. Morozov, M. Olshanetsky, A. Marshakov and
S. Shatashvili, ``Wess-Zumino-Witten Model as a Theory of
 Free Fields,'' Int. J. Mod. Phys. {\bf A5} (1990) 2495.}
\ref\bars{I. Bars, ``Free Fields and New Cosets of Current
Algebras,'' Phys. Lett. {\bf B255} (1991) 353.}
in the search of
$WZW$ cosets that have a quadratic stress-tensor upon 
bosonisation.
There is no known geometric interpretation of this coset
construction although we will attempt to formulate it
as geometrical as possible. We use the results of
\ref\df{J. de Boer and L. Feher,
``Wakimoto Realizations of Current Algebras: An Explicit Construction,''
Commun. Math. Phys. {\bf 189} (1997) 759; hep-th/9611083.}
in order to write the classical Lagrangian for
such ``cosets'. For the particular cases of
$SL(3)/SL(2)$ and $SL(4)/SP(2)$
we will see the emergence of the $AdS_5$ metric.

In section~4 we study NSR strings in these backgrounds. We describe the
bosonization including the NSR fermions and show that $N=1$
world-sheet supersymmetry is preserved.

In section~5 we combine different NSR ``cosets'' to construct critical
superstring backgrounds with $\hat{c}=10$. We discuss the
target space supersymmetries and comment on the
GSO projection. The theory that
resembles $AdS_5 \times S^5$ preserves, quite intriguingly,
18 of the 32 spacetime supersymmetries.

In section~6 we then turn to the generalization of the ``cosets''
to supergroups, and in particular focus on the
interesting case of $SL(4|4)/SP(2)^2$.
Here we will face some difficulties which are most likely
related to something which has become a standard lore by now,
namely that the only
consistent $IIB$ string background with maximal SUSY (32-supercharges) are given
by flat space or
$AdS_5 \times S^5$ (which has
$SO(6)$ global symmetry, $N$-units of $RR$ five form flux, and
no other
fields turned on)\foot{We thank A. Strominger for pointing this out to
us.}.
Our backgrounds have parity violating $WZ$
terms, a reduced global symmetry and therefore we do
not expect to be able to recover the usual $AdS_5 \times S^5$
background, although we do get a CFT on the
world-sheet whose target space metric is that of $AdS_5$.

We note that in the study of \bvw\ there are two
distinguished points in parameter space: 1. the $WZW$ point -
when the parameters in front of the kinetic and WZ terms are
related, leading to a $PSL(2|2)$ current algebra symmetry, 2. when
the $WZ$ term is absent. The sigma models studied in this paper
that are based on supergroups are analogues of 1. In the case
considered in \bvw\ the $RR$ flux is zero at this point. It is
not clear whether this is true for the higher dimensional
$AdS$ spaces. Another interesting question
that is not discussed in this paper
is whether it is in general possible
to perturb away from the WZW point, and if so, how.

 Finally, section~7 describes various open problems and directions
for future research.

\newsec{Sigma Models for $AdS_{2d+1}$}

The simplest way to introduce the conformal field theories presented
in the introduction is to start with the 
well-known case of the $SL(2,R)$ WZW model.
The conformal sigma model with $AdS_3$ target space is given by the 
WZW action for 
$SL(2,R)$ in the Gauss parametrization and 
reads \ref\as{A. Alekseev, S. Shatashvili, preprint LOMI-E-16-88, 
Nucl. Phys. {\bf B323} (1989) 719}
\eqn\adsthree{S={k \over 2\pi}
\int\,d^2 z\,(\partial\phi\bar{\partial}\phi+e^{2\phi}\partial
\bar{\gamma} \bar{\partial} \gamma) .}
This theory has an $sl_2$ current algebra.
By introducing auxiliary fields $\beta,\bar{\beta}$ and
rescaling $\phi$ the action
can be rewritten as
\eqn\adstff{S = {1 \over 4 \pi} \int \,
d^2 z \, (\partial  \phi \bar{\partial} \phi + \beta \bar{\partial}
\gamma + \bar{\beta} \partial \bar{\gamma} -
\beta \bar{\beta} e^{-2\phi/\alpha_+} -
{2\over \alpha_+} \phi \sqrt{g} R)}
where $\alpha_+ = \sqrt{2(k-2)}$. This latter action describes
a free field $\phi$ with some background charge and two free
$\beta,\gamma$ systems, perturbed by the exactly marginal
operator $V=-\beta \bar{\beta} e^{-2\phi/\alpha_+}$. The operator
$V$ is of the form $S\bar{S}$, with $S=\beta e^{-2\phi/\alpha_+}$
a dimension $(1,0)$ operator known as the screening current.
The contour integral of $S$ is known as the screening charge of
the theory. The only holomorphic operators in the theory that
commute with $\oint S$ are those constructed out of the $sl_2$
currents. In addition, the correlation functions of the theory
can, after an appropriate number of screening charges have been
inserted, be computed in the free field approximation.
In fact, the results obtained a decade ago in the study of
correlation functions for WZW models using the bosonisation technique
for conformal blocks produced results that were only valid
for compact groups; note that for $SU(2)$ what one does is
the analytic continuation $\phi \rightarrow i\phi$ -
this leads to the correct correlators although
the analytically continued sigma model doesn't describe
anything close to the $SU(2)$ WZW action, see for example
\gmmos\ for the Lagrangian approach.

To write down sigma models for $AdS_{2d+1}$, we simply take $d$
copies of the single fields $\gamma,\bar{\gamma},\beta,\bar{\beta}$,
and write down the same actions as above. The background charge in
\adstff\ remains unchanged in order for the perturbations to
be exactly marginal. However, if we integrate out the fields
$\beta,\bar{\beta}$ we now generate a nonzero background charge in
\adsthree , because each pair of $\beta,\bar{\beta}$ contributes
$+2/\alpha_+$ to the background charge. Thus, the appropriate
generalizations of \adsthree\ and \adstff\ read
\eqn\adsd{S={k \over 2\pi}
\int\,d^2 z\,(\partial\phi\bar{\partial}\phi+e^{2\phi}\partial
\bar{\gamma}_r \bar{\partial} \gamma_r)
+{1 \over 2 \pi} \int \, d^2 z \, (d-1) \phi \sqrt{g} R
.}
and
\eqn\adsdd{S = {1 \over 4 \pi} \int \,
d^2 z \, (\partial  \phi \bar{\partial} \phi + \beta_r \bar{\partial}
\gamma_r + \bar{\beta}_r \partial \bar{\gamma}_r -
\beta_r \bar{\beta}_r e^{-2\phi/\alpha_+} -
{2\over \alpha_+} \phi \sqrt{g} R)}
where $r=1,\ldots d$ is summed over, and $\alpha_+=\sqrt{2(k-2d)}$.

In \adsd\ we recognize a standard sigma model on $AdS_{2d+1}$
with nonzero $B$ field and linear dilaton. It may appear somewhat surprising
that these background fields solve (the lowest order in $\alpha'$)
supergravity equations of
motion. We will now show that this is indeed the case.

Recall that supergravity equations of motion
with only a nonzero metric, NS two-form and dilaton read
\eqn\check{
R_{ij} + 1/4 H_{ikl} H_j{}^{kl} - 2 \nabla_i \nabla_j \Phi  =  0}
\eqn\checkk{
\nabla^k H_{kij} - 2 \nabla^k \Phi H_{kij}  =  0 .
}
 For the metric we take the $AdS_{2d+1}$ metric
\eqn\metric{
ds^2 = d\phi^2 + e^{2 \phi}(d \gamma_r d \bar{\gamma}_r )
}
and for the two form
\eqn\form{
B= e^{2\phi} d \gamma_r  \wedge d\bar{\gamma}_r  ,
}
where $r=1,\ldots,d$. In addition, we take the linear dilaton from \adsd ,
\eqn\dilaton{\Phi=(d-1) \phi .}
It is now straightforward to compute
\eqn\curvature{R_{ij}=2dg_{ij}}
and since $H_{\phi\gamma_r\bar{\gamma}_r}=e^{2\phi}$, we get
\eqn\hsquare{H_{\phi kl} H_{\phi}{}^{kl} = -8dg_{\phi\phi} ,
\qquad H_{\gamma_r k l}H_{\bar{\gamma}_r}{}^{kl} = -8g_{\gamma_r\bar{\gamma}_r}. }
 Furthermore, we find that
\eqn\dila{\nabla_{\gamma_r}\nabla_{\bar{\gamma}_r} \Phi = (d-1)
 g_{\gamma_r\bar{\gamma}_r}, \qquad
\nabla^k H_{k\gamma_r\bar{\gamma}_r} = 2 (d-1) e^{2\phi}.}
Thus, both supergravity equations of motion are satisfied.

The one-loop correction to the central charge of the theory is given by
\eqn\central{c=(2d+1) + 3 \alpha' (
4 \nabla_i \Phi \nabla^i \Phi - 4 \nabla^2 \Phi + R + {1 \over 12}
H_{ijk} H^{ijk} ) }
and we find
\eqn\centrall{c  = (2d+1) + 12 \alpha'  + {\cal O}(\alpha'^2) .  }

The usual coefficient in front of \adsd\ is $(4\pi \alpha')^{-1}$,
so that $\alpha' \sim (2k)^{-1}$. We can also compute the central
charge to all orders using the free field representation \adsdd ,
with the result
\eqn\centrala{c= (2d+1) + {6 \over k-2 d} }
which agrees with \centrall\ up to terms of order $k^{-2}$.

\subsec{holographic aspects}

If we approach the boundary of $AdS_{2d+1}$, the description in terms
of free fields given by \adsdd\ becomes very useful. Both the
string coupling and the perturbation become very small near the
boundary. On the other hand, it seems that the space-time theory
is strongly coupled near the boundary, because the string coupling
goes to infinity. However, the growth of the strength of the
string interactions has to compete against the rate at which 
points at fixed $\gamma,\bar{\gamma}$ separate near the boundary. 
The situation has some similarities to the one discussed in
\ref\abks{O. Aharony, M. Berkooz, D. Kutasov and N. Seiberg,
``Linear Dilatons, NS5-Branes and Holography,'' JHEP {\bf 9810} (1998)
004, hep-th/9808149.}. To analyze these competing effects, it
is better to pass to the Einstein frame. In the Einstein frame, the metric
on $AdS_{2d+1}$ becomes
\eqn\eins{ds^2_E = e^{2\phi/(2d-1)}(e^{-2\phi} d\phi^2 +
d\gamma_r d\bar{\gamma}_r).}
Because of the positive power of $e^{\phi}$ in front of this
expression, distances on the boundary parametrized by
$\gamma_r, \bar{\gamma}_r$ go to infinity as $\phi$
goes to infinity,  and therefore we expect holographic
behavior for large $\phi$ as in \abks.

A useful set of vertex operators in the theory is given by
wave functions which are solutions
of the Laplace equation on $AdS_{2d+1}$ that behave as a delta
function near a given boundary point. As in \bort, \ks\
correlation functions of such operators will compute correlation
functions of the (unknown) theory living on the boundary.

\subsec{reality conditions}

So far we have not yet specified which reality conditions the fields
satisfy. This depends on the precise $AdS$ space we wish to consider.
Euclidean $AdS_{2d+1}$ is given by the equation
\eqn\ja{\sum_{r=1}^d (x_r^2 + y_r^2) + z^2 - w^2 =-1}
in the space with metric
\eqn\jb{ds^2 =
\sum_{r=1}^d (dx_r^2 + dy_r^2) + dz^2 - dw^2 . }
We parametrize solutions to \ja\ as follows
\eqn\jc{x_r+iy_r=\gamma_r e^{\phi}, \quad
x_r-iy_r=\bar{\gamma}_r e^{\phi},}
\eqn\jd{z+w= e^{-\phi} + e^{\phi} \sum_r \gamma_r \bar{\gamma}_r,\quad
 z-w=-e^{-\phi} .}
If we insert this into \jb\ we obtain the $AdS_{2d+1}$ metric
in the form $ds^2 = d\phi^2 + e^{2\phi} d\gamma_r d\bar{\gamma}_r$.
Thus, to describe Euclidean $AdS$ we require that $\gamma_r$
and $\bar{\gamma}_r$ are each others complex conjugate.
Other signature $AdS$ spaces are obtained by taking $\gamma_r$
and $\bar{\gamma}_r$ real and independent for certain $r$,
or by taking them to be minus each others complex conjugate.
Lorentzian signature $AdS$ has one pair of real independent
$\gamma$, $\bar{\gamma}$. All possible signature $AdS$ spaces
can be obtained this way, except $S^{2d+1}$.

It is at present not clear to us whether there exists a
generalization of the $AdS_{2d+1}$ sigma models to $S^{2d+1}$ for
$d>1$. For $d=1$, as we already mentioned, there is a close relation between
the $Sl(2,R)$ WZW theory at negative level and
the $SU(2)=S^3$ WZW theory at positive level which in terms of the variables in
\adsthree\ requires analytic continuation in $\phi$. Perhaps a similar
analytic continuation of $AdS_{2d+1}$ to negative level
has an interpretation as a CFT on $S^{2d+1}$.  Because we don't have
full understanding in the simplest case of $Sl(2,R)
\leftrightarrow SU(2)$, but analytic continuation does give
correct results,
we will assume that a
similar procedure works for the $AdS_{2d+1} \leftrightarrow
 S^{2d+1}$ case as well. It would
be interesting to investigate this further.

\subsec{global and local symmetries}

The global symmetries of \adsd , ignoring the dilaton,
are for $d>1$ given by
\eqn\glsy{\delta\phi=c,\quad
\delta \gamma_r = a_r + b_{rs} \gamma_s,\quad
\delta \bar{\gamma}_r = \bar{a}_r + \bar{b}_{rs} \bar{\gamma}_s}
with the parameters subject to the condition
\eqn\symcond{b_{rs} + \bar{b}_{sr} + 2 c \delta_{rs} = 0 .}
The reality conditions on the parameters follow from those
on the fields. The number of global symmetries is equal
to $(d+1)^2$. For Euclidean $AdS_{2d+1}$ the group of global
symmetries contains an obvious $U(d)$ group of global
symmetries. The full group is
smaller than the isometry group of
$AdS_{2d+1}$ which is $SO(2d+1,1)$. The group of global symmetries
is broken by the presence of the NS two-form, whose field
strength in terms of the coordinates \ja\ is proportional
to $H \sim (z-w)^{-1} d(z-w) \wedge dx_r \wedge dy_r$.

 For $d=1$ the global symmetries of Euclidean $AdS_3$ are
$sl(2,{\bf C})$, for Lorentzian $AdS_3$ $sl(2,{\bf R})
\times sl(2,{\bf R})$.

Besides the global symmetries there are also holomorphic
symmetries of $AdS_{2d+1}$, that give rise to holomorphic
currents. These currents are given by
\eqn\holocur{ J_r (z) = e^{2\phi} \partial \bar{\gamma}_r = \beta_r, \quad
J_0 (z)= - \partial \phi + e^{2\phi}\sum_r \gamma_r
\partial \bar{\gamma}_r = \sum_r \beta_r\gamma_r - \alpha_+\partial\phi }
and remain holomorphic in the full quantum theory (via
the
second equality above).
These operators commute with the screening charges:
\eqn\charges{Q_r=\int dz \beta_r e^{-\frac{2}{\alpha_+}\phi}}
and also form the OPE algebra:
\eqn\alg{\eqalign{& J_r(z)J_{r'}(w) = 0, \quad J_r(z)J_0(w) =
\frac{-2J_r(z)}{z-w}\cr
& J_0(z)J_0(w) = \frac{-2k}{(z-w)^2}}}
In order to determine the complete chiral algebra one should
find all
polynomials in the free fields and their derivatives that commute with
the screening charges. One such operator is the stress-tensor:
\eqn\stress{T(z) =-\frac{1}{2} (
 \sum_r \beta_r \partial \gamma_r +
 (\partial \phi)^2 + \frac{2}{\alpha_+}\partial^2\phi ) .}

\newsec{Bosonic WZW cosets}

It was noticed in \ff, \gmmos, \bars\ that given a group $G$ and a subgroup
$H=H_1 \times H_2 \times ...$, such that the levels of the corresponding
current algebras satisfy
 $k^G+c^G_V = k^{H_1} + c_V^{H_1}
= ...$, one can attempt to define some sort of ``coset'' theory,
by getting rid of the degrees of freedom residing in $H$.
The basic
idea of this construction is to do a partial bosonization of
the $G$ currents, by expressing them in terms
of the $H$ currents and extra $\beta, \gamma, \phi$
systems,
and to subsequently set the $H$ currents
to zero. One finds that the stress tensor that survives is
quadratic in the remaining fields. For a special value of the
level $k^G$, the original $G$ current algebra survives.
General expressions for the partial bosonizations that one
needs to implement this construction are given in \df.
We first illustrate this procedure for the case
$SL(3)/SL(2)$ in the Lagrangian formalism, and then turn
to more general examples.

\subsec{SL(3)/SL(2)}

The discussion in this section will be completely classical.
To discuss the coset for $SL(3)/SL(2)$, we start by taking the
following Gauss decomposition for $SL(3)$
\eqn\gaus{
G= \pmatrix{1 & \bar{\gamma}_1 & \bar{\gamma}_2 \cr
                   0 & 1 & 0 \cr
 0 & 0 & 1  \cr}
 g_{sl_2}
\pmatrix{ e^{-2 \phi} & 0 & 0 \cr
                   0 & e^{\phi} & 0 \cr
                   0 & 0 &  e^{\phi} \cr}
\pmatrix{1 & 0 & 0 \cr
                   \gamma_1 & 1 & 0 \cr
                   \gamma_2 & 0 & 1 \cr}
}
where $g_{sl_2}$ is a group element of sl(2), which
is embedded in $sl_3$ in the bottom right $2\times 2$ block. We write
$G=g_U g_{sl_2} g_D g_L$. The WZW action evaluated for $G$ is
\eqn\action{
S_{WZW}(G) = S_{WZW}(g_D) + S_{WZW}(g_{sl_2}) +
{k \over 2\pi}
 \int {\rm tr} (g_U^{-1}
\partial g_U
g_{sl_2} g_D \bar{\partial} g_L g_L^{-1} g_D^{-1} g_{sl_2}^{-1})
}
The last term above can be written explicitly:
\eqn\imp{e^{3\phi}\partial \bar{\gamma}_i g_{sl_2}^{ij}\bar \partial \gamma_j .}
This form shows the interaction
between $SL(2)$ and remaining degrees of freedom. The idea of the
novel type of coset is to set $g_{sl_2}=1$. Classically this is
not problematic, but in the quantum theory $g_{sl_2}$ has a nonzero
conformal weight and cannot simply be put equal to one.
Later we will see how the theory has to be modified so as to
preserve conformal invariance after $g_{sl_2}$ has been put equal to one.
After putting $g_{sl_2}$ equal to one, the target space $M$ of the
theory will be the submanifold $g_Ug_Dg_L \subset G$ parametrized
by $g_U$, $g_D$ and $g_L$. This is not quite the homogeneous
space $SL(3)/SL(2)$, because tangent vectors to $M$ are not 
orthogonal to the $sl_2$ Killing vectors, and the corresponding
``Faddeev-Popov'' determinant is absent. It is also not
the target space of the usual coset construction. In fact, as we will
see below, it is $AdS_5$.

The next step is to bosonize \action\ in the Lagrangian
approach, following \gmmos, which involves changing
variables from $\bar{\gamma}_i$ and $\gamma_i$ to $\beta_i$
and $\gamma_i$, so that the last term in \action\
becomes the free action $\int \beta_i \bar \partial \gamma_i$.
Therefore, the change of variables reads
\eqn\change{
\beta_i = e^{3\phi}g_{sl_2}\partial \bar{\gamma}_i .}
This change of variables has a Jacobian which can be found for
example in \gmmos. This Jacobian leads to a shift in the coefficient of
the kinetic term for $\phi$, introduces an improvement term for
$\phi$ (i.e. a linear
dilaton background) and changes the level of the $sl_2$ WZW theory,
all in such a way that the conformal invariance and central
charge of the theory remain unaltered. However, the discussion in
this section will be purely classical and we will ignore this
Jacobian.

In matrix form, the change of variables \change\ reads
\eqn\free{ \pmatrix{ 0 & \beta_1 & \beta_2 \cr
                   0 & 0 & 0 \cr
 0 & 0 & 0 \cr}
= g_D^{-1} g_{sl_2}^{-1} g_U^{-1} \partial g_U
  g_{sl_2} g_D
}

When inserted in the action we find that the last term becomes the sum of
two free beta-gamma systems, and it is also immediately clear that
the stress-energy tensor will be the direct sum of that of the WZW
theories for $g_D$,
$g_{sl_2}$ and that of the two beta-gamma systems. By working out the
$sl_3$ current $G^{-1} \partial G$, we find an expression for the
$sl_3$ current in terms of the $sl_2$ currents, and
$\gamma_1,\gamma_2,\beta_1,\beta_2,\phi$. This is the realization as in (4.7.22)
of \gmmos.
In addition, \free\ enables us to express $\partial \gamma_1$
and $\partial \gamma_2$ in terms of the other fields. Requiring that
the integral of these vanish, or are equal
to the monodromies of $\gamma$, provides the two screening charges for this
type of free field realization. (The number of screening 
charges is always
$\dim g_L= \dim g_U$.)
Now let us
try to remove the $sl_2$ degrees of freedom by putting $g_{sl_2}=1$.
The relation \free
with $g_{sl_2}=1$ becomes
\eqn\freeone{ \pmatrix{ 0 & \beta_1 & \beta_2 \cr
                   0 & 0 & 0 \cr
 0 & 0 & 0 \cr}
= g_D^{-1} g_U^{-1} \partial g_U
  g_D
= \pmatrix{ 0 & e^{3\phi} \partial \bar{\gamma}_1 & e^{3\phi}
                          \partial \bar{\gamma}_2 \cr
                   0 & 0 & 0 \cr
 0 & 0 & 0 \cr}
}
The screening charges
 are therefore $\int e^{-3 \phi} \beta_1$ and $\int e^{-3 \phi}
\beta_2$,
similar to (4.1.22-24) of \gmmos\ with the $sl_2$ degrees of freedom removed.
We now want to use these screening charges to find
the action of the reduced
theory.
Thus we substitute $\beta_1=e^{3\phi}
 \partial \bar{\gamma}_1$ and $\beta_2=e^{3\phi}
\partial \bar{\gamma}_2$ back into the action
\eqn\actiontwo{
S=S_{WZW}(g_D) + {k \over 2\pi}
\int (\beta_1 \bar{\partial} \gamma_1 + \beta_2 \bar{\partial}
\gamma_2)
}
and get
\eqn\actionthree{
S={1\over 2\pi} \int (3 \partial \phi \bar{\partial} \phi + e^{3\phi} (
 \partial \bar{\gamma}_1 \bar{\partial} \gamma_1+
 \partial \bar{\gamma}_2 \bar{\partial} \gamma_2 ))
}
Up to a rescaling, this is the action of $AdS_5$ that we presented
in the previous section.
The final action in terms of group variables simply reads
\eqn\simp{
S=S_{WZW}(g_U g_D g_L). }
Thus the process of first bosonizing, removing the $sl_2$ degrees of freedom
and then undoing the bosonization amounts to the same as removing
the $sl_2$ degrees of freedom from the Gauss decomposition of the $SL(3)$ group
element. In fact, as will we argue later, the action \simp\ describes
an exact conformal field theory once a suitable linear dilaton background,
which makes the conformal weights of all screening currents
equal the one, is included.

\subsec{Bosonization}

Let us now discuss the bosonization of the sigma model for
$AdS_{2d+1}$, given in \adsd , in the path integral framework.
The action reads
\eqn\aadsd{S={k \over 2\pi}
\int\,d^2 z\,(\partial\phi\bar{\partial}\phi+e^{2\phi}\partial
\bar{\gamma}_r \bar{\partial} \gamma_r)
+{1 \over 2 \pi} \int \, d^2 z \, (d-1) \phi \sqrt{g} R
.}

The measure
in the path integral is defined through the sigma model metric as
\eqn\meas{\int [(\delta\phi)^2 + e^{2\phi}\sum \delta\gamma_r
\delta \bar{\gamma}_r]}

We can map the sigma model to a theory of
free fields if we introduce new fields
\eqn\change{\beta_r= e^{2\phi}\partial \bar{\gamma}_r}
The Jacobian $J$ for this change of variables is nontrivial and can be
computed through the anomaly
in the determinant
\eqn\determ{J = det^d[e^{-2\phi}\bar \partial e^{2\phi} \partial]}
exactly like in \gmmos\ leading to the result
\eqn\answ{log J = d/(4\pi)\int d^2 z\, [4 \partial \phi
\bar \partial \phi + 2 R \phi]}
At the same time after the above change of variables the measure becomes the
standard one for the free fields
$\beta_r, \gamma_r, \phi$ and finally we get the quantum action
\eqn\aadsdd{S_q = {1 \over 4 \pi} \int \,
d^2 z \, ((2k-4d)\partial  \phi \bar{\partial} \phi + 2 k \beta_r \bar{\partial}
\gamma_r  - 2
\phi \sqrt{g} R).}

The change of variables also produces the screening charges
\eqn\scr{\int \partial \bar{\gamma}_r = \int \beta_r e^{-2\phi} . }
The background charge of the field $\phi$ in \aadsdd\ is precisely
such that the screening charge has conformal weight zero, which is
a necessary condition for conformal invariance.

At this stage we have obtained two bosonized version of the sigma model \adsd,
namely the chirally bosonized version \aadsdd\ and the non-chirally
bosonized version \adsdd . One of the main applications of free
field realization for WZW models is that it allows one to compute correlator functions.
This has been worked out in \ff, \gmmos,
\ref\zam{A.B. Zamolodchikov, unpublished.}, \ref\ffa{D.~Bernard, G.~Felder,
``Fock representations and BRST cohomology in $SL(2)$ current algebra,''
Commun.\ Math.\ Phys.\ {\bf 127} (1990) 145.},
\ref\gw{K. Gawedzki, ``Quadrature of Conformal Field Theory,''
Nucl. Phys. {\bf B328} (1989) 733.}
\ref\ffb{P.~Bouwknegt, J.~McCarthy, K.~Pilch,
``Free field realizations of WZNW models, the BRST complex and its
quantum group structure,''
Phys.\  Lett.\  {\bf B234} (1990) 297;
``Quantum group structure in Fock space resolutions of $\widehat{sl}(n)$
representations,''
Commun.\ Math.\ Phys.\ {\bf 131} (1990) 125;
``Free field approach to two-dimensional conformal field theory,''
Prog.\ Theor.\ Phys.\ Suppl.\ {\bf 102} (1990) 67;
``Some aspects of free field resolutions in 2D CFT with applications to
quantum Drinfeld-Sokolov reduction,''
p.~407 in:  Strings and Symmetries 1991, eds.\  N.~Berkovits et al.,
World Scientific, Singapore, 1992, hep-th/9110007.},
\ref\ffc{H.~Awata, A.~Tsuchiya, Y.~Yamada,
``Integral formulas for the WZNW correlation functions,''
Nucl.\ Phys.\ {\bf B365} (1991) 680;
H.~Awata,
``Screening currents Ward identity and integral formulas
for WZNW correlation functions,''
Prog.\ Theor.\ Phys.\ Suppl.\ {\bf  110} (1992) 303,
hep-th/9202032.}
\ref\ffd{
G.~Kuroki,
``Fock space representations of affine Lie algebras and integral
representations in the Wess-Zumino-Witten models,''
Commun.\ Math.\ Phys.\ {\bf 142} (1991) 511.} .

The chiral bosonized version of the action can be used to compute the
conformal blocks of the CFT. To combine left and right-moving conformal
blocks we need the non-chirally bosonized action.
The non-chirally bosonized action consists of the chiral and anti-chiral
bosonized action, perturbed by operators of the form $S\bar{S}$ where
$S$ is often a screening current. Suppose that we are interested in studying
a CFT with some chiral algebra $A$. If $A$ is generated by the operators
that commute with a given set of screening charges $Q_r=\oint S_r$,
then the conformal blocks of the theory can be studied using only the
screening charges $Q_r$, which are called the simple screening charges.
One might think that the action describing the full CFT is then
simply the sum of the chiral and anti-chiral bosonized action perturbed by
$\sum S_r \bar{S}_r$. This is not the full story however. The variation
of $S_r$ under the transformations generated by an element of $A$ is
a total derivative, $\delta S_r = \partial Y_r$, so that the screening
charges $Q_r$ are invariant. It is then clear that the variation of
$\int d^2 z \, \sum_r S_r \bar{S}_r$ is not zero but rather
$-\int d^2 z \, \sum_r Y_r \partial \bar{S}_r$. This vanishes on-shell
but not off-shell. To find an action which is invariant, we need to
modify the transformation rules of the fields and sometimes include
additional terms in the action that look like higher powers
of the simple screening currents, and which give rise to the non-simple
screening charges.
 This does not happen in \adsdd, but if
we bosonize the $SL(3)$ WZW theory we find that the free field action
perturbed by the two simple screening currents
 does not have $SL(3)$ current algebra;
for that one needs to add an additional term to the action,
which is built out of iterated simple screening currents, i.e.
non-simple screening currents.
After the addition of the additional term, the action is identical
to the classical $SL(3)$ WZW action. This procedure is certainly
correct for $SL(N)$, but we are not aware of any good explanation in
the literature. (As we mentioned before, bosonization allows one
to exactly solve WZW theories with compact groups like $SU(N)$,
by analytic continuation of the results for $SL(N)$. The
precise meaning of holomorphic factorization for noncompact
groups like $SL(2,R)$ is not clear; one expect effects similar
to the Liouville model, where one finds that correlators depend
non-analytically on the cosmological constant $\mu$, which
is the coefficient in front of the perturbation $e^{\phi}$
in the Lagrangian.)
The point we
want to make is that in order to compute certain quantities it is sufficient
to give a set of simple screening charges
 and the symmetries of the problem one
wants to preserve. The simple screening charges
 give the conformal blocks, and
the coupling of left and right movers follows by imposing the symmetries
of the problem. The symmetries of the problem allow one to construct,
order by order, the action that is invariant under those symmetries, which
then in turns provides the required coupling of left and right movers.
An example of this will appear in section~4 of the paper.

\subsec{General Bosonic Case}

The type of coset construction we have been considering can in principle
be done for any generalized Gauss decomposition $g=g_u h_0 g_d g_l$.
Here, $h_0$ takes values in a product of simple factors $h_0 \in H \equiv
H_1 \times H_2 \times \ldots$, $g_d$ takes values in an abelian group,
and $g_d$ and $g_l$ are certain upper and lower triangular matrices.
The idea of the coset construction is to somehow get rid of the
degrees of freedom residing in $H$ \gmmos .

In case the subgroup $G_u H G_d$ of $G$ is parabolic, the construction
of \df\ can be used to find an explicit representation of the $G$ currents
in terms of $\beta,\gamma$ systems associated to $G_{u,l}$, scalar
fields $\phi_i$ parametrizing $G_d$, and $H=H_1 \times H_2 \ldots
$ currents. The level $k_i$ of the
$H_i$ currents is $k_i + c_V^{H_i} = k^G + c_V^{G}$. In addition, there
is a set of screening charges, and the stress tensor of the theory
contains improvement terms for the scalars $\phi_i$.
To get rid of the degrees of freedom in $H$ we cannot use conventional
coset techniques, because the $H$ currents themselves are not part
of the theory (they do not commute with the screening charges). 
This is reflected
in the fact that the level of the $H_i$ current algebra differs from
that of $G$. Therefore we propose to do the following: remove the $H$
degrees of freedom by hand, and subsequently adjust the background
charges in such a way that the screening charges keep the right conformal
weight. Classically this amounts to replacing $S_{\rm WZW}(g_u h_0 g_d
g_l)$
by $S_{\rm WZW}(g_u g_d g_l)$, but in the quantum theory we will also
find a nontrivial dilaton. Because the screening charges give exactly
marginal deformations of the reduced theory, this procedure
will always give an exact conformal background. Its correlation functions
can in principle be computed using free field methods. Nevertheless,
it would be interesting to know if there is a more direct way to obtain
the reduced theory from the original theory. For instance, one could
try to gauge the $H$-degrees of freedom, thereby replacing the $H$-WZW
theory by a topological $H/H$ model, or one could try to define
the reduced theory as the BRST cohomology of a suitable BRST operator
of the original theory.

We illustrate the general method now by considering the case $G=SL(N)$
and $H=SL(p_1) \times \ldots \times SL(p_r)$, with $\sum p_i = N$.
The $SL(p_i)$ are embedded as diagonal $p_i \times p_i$ matrices in
$SL(N)$, with $SL(p_1)$ appearing in the top left block, and
$SL(p_r)$ appearing in the right bottom block. The abelian group
$G_d$ is parametrized by $r-1$ scalar fields $\phi_i$. In addition there
are
$r-1$ screening charges, associated to the simple roots
$\alpha_{p_1},\alpha_{p_1+p_2},\ldots,\alpha_{p_1+\ldots+p_{r-1}}$.
The $i^{\rm th}$ screening charge is composed out of polynomials in
the $\beta,\gamma$-fields, a vertex operator for the $H$-theory
that creates the $p_i p_{i+1}$ dimensional (bifundamental)
representation of $SL(p_i) \times SL(p_{i+1})$, and an exponential
of the scalar fields $\phi_i$. If we now remove the $H$ degrees
of freedom from the theory, the $i^{\rm th}$ screening charge decomposes
into $p_i p_{i+1}$ screening charges, each with the same exponent of
$\phi_i$. To make sure that all screening charges are contour integrals
of objects of conformal weight one, we need to adjust the background
charges of the scalar fields $\phi_i$. Since there are $r-1$ different
exponentials of the fields $\phi_i$, we get $r-1$ equations for $r-1$
background charges, that have a unique solution. The stress tensor of
the reduced theory is then the sum of free $\beta,\gamma$ stress tensors
and improved stress tensors for $\phi_i$. The central charge of
this theory can be shown to be equal to
\eqn\cen{c= {\rm dim} \, G_d + {\rm dim} \, G_u + {\rm dim} \,
G_l + { 3 y \over k - c_V^G}}
where
\eqn\defy{y = \sum_{i=1}^{r-1}
({1 \over p_i} + {1 \over p_{i+1}} )^2
{ (p_{i+1} + \ldots p_{N})(p_1 + \ldots + p_i) \over N}\qquad }

\eqn\auxc{ \qquad\qquad  + 2 \sum_{1 \leq i < j \leq r-1}
({1 \over p_i} + {1 \over p_{i+1}} )
({1 \over p_j} + {1 \over p_{j+1}} )
{ (p_{j+1} + \ldots p_{N})(p_1 + \ldots + p_i) \over N}  .  }

As a check, we notice that for $p_i=1$ and $r=N$, we get $y=N(N^2-1)/3$
and we recover from \cen\ the usual central charge
of the $SL(N)$ WZW theory. In this case the reduced theory is equivalent
to the original WZW theory, as $H$ is trivial.

Another example that is straightforward is $SL(d+1)/SL(d)$. This
corresponds
to $r=2$ and $p_1=1,p_2=d$. We get $y=(d+1)/d$ and
\eqn\auxa{c=(2d+1) + {3(d+1) \over d(k-d-1)} .   }
The resulting theory
describes strings propagating on $AdS_{2d+1}$. To see this,
we generalize the example described in the beginning of section~2.1
to $SL(d+1)$, by taking $G_d={\rm diag}(e^{-d\phi},e^{\phi},
\ldots,e^{\phi})$. We obtain the classical action
\eqn\auxd{S={k \over 2 \pi} \int d^2 z (d(d+1) \partial \phi
\bar{\partial}{\phi} + 2 e^{(d+1)\phi} \sum_r \partial \bar{\gamma}_r
\bar{\partial} \gamma_r ).}
After an appropriate rescaling of the fields we obtain the action as
in \adsd\ but with $k$ replaced by $2kd/(d+1)$. If we make the same
replacement for $k$ in \centrala, we obtain precisely \auxa,
as expected.

\newsec{NSR Formulation}

We now turn to the supersymmetrization of the $AdS_{2d+1}$
sigma models of section~2, so that they will have $N=1$
world-sheet supersymmetry. To achieve this, we write the
sigma models of section~2 in terms of the following superfields
\eqn\fields{\eqalign{
\Phi = \phi + \theta \lambda^L + \bar{\theta} \lambda^R +
 \theta \bar{\theta} F \cr
\Gamma_i = \gamma_i + \theta \psi^L_i + \bar{\theta} \psi^R_i +
\theta \bar{\theta} G_i \cr
\bar{\Gamma}_i = \bar{\gamma}_i + \theta \bar{\psi}^L_i
+ \bar{\theta} \bar{\psi}^R_i +
\theta \bar{\theta} \bar{G}_i \cr
S_i = \sigma_i + \theta \beta_i + \bar{\theta} u_i
 + \theta \bar{\theta} H_i \cr
\bar{S}_i = \bar{\sigma}_i + \theta \bar{u}_i + \bar{\theta} \bar{\beta}_i
 + \theta \bar{\theta} \bar{H}_i }}
The supersymmetric versions of \adsd\ and \adsdd\ read
\eqn\act{
S={\alpha_+^2 \over 4\pi} \int d^2 z d^2 \theta (D_+ \Phi D_- \Phi +
e^{2\Phi} D_+ \bar{\Gamma}_i D_- \Gamma_i)}
and
\eqn\actt{
S={\alpha_+^2 \over 4\pi}
\int d^2 z d^2 \theta (D_+ \Phi D_- \Phi - \bar{S}_i D_+ \bar{\Gamma_i}
+ S_i D_- \Gamma_i -
e^{-2\Phi} S_i \bar{S}_i) }
where $D_+ = \partial_{\theta} + \theta \partial$ and $D_-=
\partial_{\bar{\theta}} + \bar{\theta} \bar{\partial}$.
After integrating out the superfields $S_i,\bar{S}_i$
from \actt\ we clearly
recover \act. In addition, there is a linear coupling of $\Phi$ to the
$N=1$ super world-sheet curvature 
\ref\howe{
P.S. Howe, ``Super Weyl Transformations in Two Dimensions,''
J. Phys. {\bf A12} (1979) 393;
M. Brown and S.J. Gates, ``Superspace Bianchi Identities and the
Supercovariant Derivative,'' Ann. Phys. {\bf 122} (1979) 443.}
that we have not explicitly written

Next, we study these actions in components. From action \act\ we get
\eqn\comp{\eqalign{
{\alpha_+^2 \over 4\pi} \int ( -\lambda^L \bar{\partial} \lambda^L +
\partial\phi \bar{\partial} \phi + F^2 + \partial \lambda^R \lambda^R \cr
+ e^{2\phi}(-\bar{\psi}^L_i \bar{\partial} \psi^L_i  +
\partial \bar{\gamma}_i \bar{\partial} \gamma_i - G_i \bar{G}_i
+ \partial \bar{\psi}^R_i \psi^R_i ) \cr
+ 2e^{2\phi}
\lambda^L(-\bar{G}_i \psi_i^R + \bar{\psi}^L_i \bar{\partial} \gamma_i)
\cr
+ 2e^{2\phi}\lambda^R (-\partial\bar{\gamma}_i \psi_i^R - \bar{\psi}^L_i G_i )
\cr
+ e^{2\phi} (2F-4\lambda^L \lambda^R) \bar{\psi}^L_i \psi^R_i )} }
The auxiliary fields should be eliminated. They
are given by
\eqn\aux{\eqalign{
 F = - e^{2\phi} \bar{\psi}^L_i \psi^R_i \cr
G_i = - 2\lambda^L \psi_i^R \cr
\bar{G}_i = -2 \lambda^R \bar{\psi}^L_i} }
and when plugged back into \comp\ we finally get
\eqn\compp{\eqalign{
{\alpha_+^2 \over 4\pi} \int ( -\lambda^L \bar{\partial} \lambda^L +
\partial\phi \bar{\partial} \phi + \partial \lambda^R \lambda^R \cr
+ e^{2\phi}(-\bar{\psi}^L_i \bar{\partial} \psi^L_i  +
\partial \bar{\gamma}_i \bar{\partial} \gamma_i
+ \partial \bar{\psi}^R_i \psi^R_i ) \cr
+ 2 e^{2\phi}(\lambda^L \bar{\psi}^L_i \bar{\partial} \gamma_i
-\lambda^R \partial\bar{\gamma}_i \psi_i^R )
\cr
- e^{4\phi} \bar{\psi}^L_i \psi^R_i
\bar{\psi}^L_j \psi^R_j) }}

The third line contains some terms that make the fermion kinetic
terms covariant, the last line is the four-fermion term that
multiplies the curvature of the connection with torsion.
This curvature is not supposed to be zero unless we are in
$AdS_3$. Indeed, the last line vanishes for $AdS_3$. This
is because the beta-functions state that the Ricci tensor with
torsion is equal to the second covariant derivative of the
dilaton. So the nonvanishing four fermion term is directly
related to the presence of a nonzero dilaton.


To bosonize the action \compp\ and to write it as a theory
of free fields perturbed by exactly marginal operators,
it is easier to start with \actt\ rather than trying to
bosonize \compp\ directly.

The component expansion of \actt\ reads
\eqn\coo{\eqalign{
{\alpha_+^2 \over 4\pi}
\int ( -\lambda^L \bar{\partial} \lambda^L +
\partial\phi \bar{\partial} \phi + F^2 + \partial \lambda^R \lambda^R \cr
-\bar{\sigma}_i \partial \bar{\psi}^R_i - \bar{u}_i \bar{G}_i
+\bar{\beta}_i \partial \bar{\gamma}_i  -
\bar{H}_i \bar{\psi}^L_i \cr
-\sigma_i \bar{\partial} \psi^L_i +
\beta_i \bar{\partial} \gamma_i - u_i G_i + H_i \psi_i^R \cr
-e^{-2\phi}(\sigma_i H_i + \beta_i \bar{\beta}_i - u_i \bar{u}_i
 + H_i \bar{\sigma}_i) \cr
-2e^{-2\phi} \lambda^L( u_i \bar{\sigma}_i - \sigma_i \bar{\beta}_i) \cr
+2 e^{-2\phi} \lambda^R (\beta_i \bar{\sigma}_i - \sigma_i \bar{u}_i ) \cr
+ e^{-2\phi} (2 F+ 4 \lambda^L \lambda^R) \sigma_i \bar{\sigma}_i ) } }

The role of $e^{2\phi}\psi^R_i$ and
$e^{2\phi}\bar{\psi}^L_i$ is taken over by $\bar{\sigma}_i$ and
$\sigma_i$ respectively. This is clear from the $H_i$ and
$\bar{H}_i$ equations of motion.
After eliminating $H,G,u$ from \coo\
and a rescaling of the fields
we get the following form
of the action
\eqn\cooo{\eqalign{
{1 \over 4\pi}
\int ( -\lambda^L \bar{\partial} \lambda^L +
\partial\phi \bar{\partial} \phi + \partial \lambda^R \lambda^R \cr
-\bar{\sigma}_i \partial \bar{\psi}^R_i
+\bar{\beta}_i \partial \bar{\gamma}_i
-\sigma_i \bar{\partial} \psi^L_i +
\beta_i \bar{\partial} \gamma_i  \cr
-e^{-2\phi/\alpha_+}(\bar{\beta}_i -{2\over \alpha_+}
 \lambda^R \bar{\sigma}_i)(\beta_i -
{2 \over \alpha_+} \lambda^L \sigma_i) \cr
-e^{-4\phi/\alpha_+} \sigma_i \bar{\sigma}_i \sigma_j \bar{\sigma}_j )
}}
This is probably the most useful form of the action.
It has the structure discussed in section~3.2. There are
$d$ simple screening charges given by
\eqn\defscr{S_i = e^{-2\phi/\alpha_+} (\beta_i - {2\over\alpha_+} \lambda^L
\sigma_i) }
but the action is not just the sum of a free field piece and
$S_i\bar{S}_i$, there is also
the term
$e^{-4\phi/\alpha_+} \sigma_i \bar{\sigma}_i \sigma_j \bar{\sigma}_j$.
As is clear from the prefactor $e^{-4\phi/\alpha_+}$, this term
is of higher order in the $e^{-2\phi/\alpha_+}$ expansion and it is
needed to restore the $N=1$ world-sheet invariance of the theory.
According to our discussion in section~3.2, it will not play a role
in the construction of the conformal blocks of the theory, but
is important in order to correctly combine left and right movers.
A further clarification of this point is certainly desirable.

As a side remark, we notice that it is possible to write everything
as free field theory perturbed by simple screening currents only. 
To accomplish this,
we first add to \cooo
\eqn\pia{{1\over 4\pi} \int d^2 z \, (A_{ij} + e^{-2\phi/\alpha_+}\sigma_i
\sigma_j)(\bar{A}_{ij} + e^{-2\phi/\alpha_+} \bar{\sigma}_i \bar{\sigma}_j),
}
which removes the four fermion terms but at the cost of
introducing $d(d-1)/2$ auxiliary fields $A_{ij}$. Next we make the change
of variables $A_{ij}=\partial \rho_{ij}$, $\bar{A}_{ij}=\bar{\partial}
\bar{\rho}_{ij}$, and represent the corresponding determinant by an
integral over $d(d-1)/2$ pairs of fermionic $\beta^{\theta}_{ij},
\theta_{ij},\bar{\beta}^{\bar{\theta}}_{ij},\bar{\theta}_{ij}$ systems.
Altogether, the four-fermion terms have now been replaced by
the sum $S_f+S_s$ of the free field action
\eqn\yaa{S_f={1\over 4\pi} \int d^2 z\,
(\beta^{\theta}_{ij} \bar{\partial} \theta_{ij} +
 \bar{\beta}^{\bar{\theta}}_{ij} \partial \bar{\theta}_{ij} +
 \partial \rho_{ij} \bar{\partial} \bar{\rho}_{ij} ) }
and the perturbation by simple screening currents
\eqn\yab{S_s ={1\over 4\pi} \int d^2 z\,
( e^{-2\phi/\alpha_+} \bar{\partial} \bar{\rho}_{ij} \sigma_i \sigma_j +
 e^{-2\phi/\alpha_+} \partial \rho_{ij} \bar{\sigma}_i \bar{\sigma}_j ) .}
It is not clear, however, what replaces the $N=1$ algebra in this
formulation of the theory, which is somewhat reminiscent of the
formulation in \bvw .

We now discuss the $N=1$ world-sheet superconformal
algebra of \cooo. The $N=1$ super stress-energy tensor
of \act\ is
\eqn\str{\alpha_+^{-2} G = \partial \Phi D_+ \Phi
- e^{2\Phi} D_+ \Phi D_+ \Gamma_i D_+ \bar{\Gamma}_i +
{1\over 2} e^{2\Phi} \partial \Gamma_i D_+ \bar{\Gamma}_i +
{1\over 2} e^{2\Phi} D_+ \Gamma_i \partial \bar{\Gamma}_i }
and that of \actt\ is
\eqn\strr{\alpha_+^{-2} G = \partial \Phi D_+ \Phi + {1 \over 2} D_+ S_i
D_+ \Gamma_i + {1 \over 2} S_i \partial \Gamma_i .}
This does not yet contain the improvement term due to the dilaton.
The improvement term is simply given by
\eqn\imr{G_{dil} = Q D_+ \partial \Phi}
Let us now focus on \strr\ and work it out in components. This will
then give $G$ and $T$ for \cooo. We get
\eqn\defg{
G=\partial \phi \lambda^L + {1\over 2} \beta_i \psi_i^L +
{1 \over 2} \sigma_i \partial \gamma_i + Q \partial \lambda^L}
\eqn\deft{
T=-{1\over 2}(\partial \lambda^L \lambda^L + \partial \phi \partial \phi +
{1\over 2} \partial \sigma_i \psi_i^L +
 \beta_i \partial \gamma_i -
{1 \over 2} \sigma_i \partial \psi_i^L + Q \partial^2 \phi) .}
These are the suitably normalized generators of the $N=1$ superconformal
algebra with central charge $c={3(2d+1) \over 2 } + 3 Q^2$.
Their form agrees with the standard form of the $N=1$ superconformal
algebra for free fields with some improvement terms.

We need to verify whether the simple screening charges
\defscr\ indeed commute with the $N=1$ algebra.
 From the action \cooo\ we obtain the following
free field OPE's
\eqn\opea{ \phi(z) \phi(w) \sim - \log |z-w|^2,
\quad \lambda^L(z) \lambda^L(w) \sim -(z-w)^{-1},}
\eqn\opeb{ \beta_i(z) \gamma_j(w) \sim -2 \delta_{ij} (z-w)^{-1} ,
\quad
\sigma_i(z) \psi^L_j(w) \sim -2 \delta_{ij} (z-w)^{-1} }
In order for the screening currents to have conformal weight one, we need
that \eqn\abcs{Q=\frac{2}{\alpha_+}.} Furthermore, the screening charges
indeed preserve the complete $N=1$
structure, because we have the OPE
\eqn\ope{e^{-2\phi/\alpha_+}(\beta_i-
{2 \over \alpha_+} \lambda^L \sigma_i)(z) \,\,\, G(w)
\sim {-
e^{-2\phi/\alpha_+}\sigma_i(w) \over (z-w)^2} + {\rm regular} }


\newsec{Spacetime Supersymmetry}

We next discuss spacetime supersymmetry of the $N=1$ supersymmetric
$AdS$ sigma models.
We take the compactification with $AdS_{2d_1+1} \times
S^{2d_2+1}$. What we mean by $S^{2d_2+1}$ is explained in
section~2.2.
The notation that
we will use is as follows: variables and fields that describe
$AdS_{2d_1+1}$ will
be denoted by the same symbols as in the previous section,
the variables and fields from the second factor $S^{2d_2+1}$
will be denoted by the same symbols as well,
but with a tilde on them. The fields and quantities with a tilde
have not yet been analytically continued.
In particular, to get the right value
for the central charge $c=15$, we need $d_1+d_2+1=5$ and
$\alpha_+ = \pm i \tilde{\alpha}_+$.
The zero modes
of the fermions in the RR sector form some kind of Clifford
algebra. We have for instance
\eqn\clif{ \{ (\sigma_i)_0,(\psi_j^L)_0 \} = - 2 \delta_{ij} ,
\quad \{ \lambda^L_0, \lambda^L_0 \} = - 1. }
Thus the zero modes can be represented via gamma matrices.
We denote by $\Gamma_{\psi}$ the gamma matrix representing
the zero mode of $\psi$. Now the RR vacua of the fermions
are created by spin fields $V_{\alpha}$, where $\alpha$
runs from $1$ to $2^{d_1+d_2+1}$. We also have the OPE
\eqn\sfope{\psi(z) V_{\alpha}(w) =
{ (\Gamma_{\psi})_{\alpha}^{\beta} V_{\beta}(w) \over
(z-w)^{1/2} } + \ldots }

An explicit
representation for the gamma matrices can be given
using matrices like
\eqn\exmat{\Gamma \sim \sigma_3 \otimes \ldots \otimes \sigma_3
\otimes \sigma_i \otimes 1 \otimes \ldots \otimes 1, \quad i=1,2 .}

As in the usual NSR string, we will attempt to construct the
spacetime supersymmetry generators using RR ground states
of conformal weight $5/8$ and the bosonized superreparametrization
ghosts.
What makes life complicated is that the most general
vertex operator that creates a RR ground state is not some
linear combination of the $V_{\alpha}$. This is because
we have the fields $\gamma_i$, whose zero modes are well-defined.
These zero modes can be multiplied arbitrarily with the
RR vacua and the state will still be a RR vacuum.
Thus the most general vertex operator that creates an RR ground
state is
\eqn\genve{V \equiv f^{\alpha}(\gamma_i,\tilde{\gamma}_i) V_{\alpha}. }

We now want to examine which vertex operators survive in the perturbed
$N=1$ theory. For this we need to check that they commute with
the screening charges, 
and that they have correct OPE's with the supersymmetry
generators.

We start by looking at the OPE with the screening currents.
The $e^{-2\phi/\alpha_+}$ factor doesn't do anything and can be ignored.
What remains is
\eqn\opeba{(\beta_i - {2\over\alpha_+} \lambda^L \sigma_i )(z) \,\,\,
V(w) \sim {
-2 {\partial \over \partial \gamma_i }f^{\alpha}(\gamma_i,
\tilde{\gamma_i} ) V_{\alpha} + {2\over \alpha_+} f^{\alpha}(\gamma_i,
\tilde{\gamma_i} )
(\Gamma_{\lambda^L} \Gamma_{\sigma_i})_{\alpha}^{\beta} V_{\beta}
\over (z-w) } + \ldots}

Thus we find a set of first order differential equations for the
functions $f^{\alpha}$. Probably, these are equivalent to the Killing
spinor equations in $AdS_{2d_1+1} \times S^{2d_2+1}$ with NS
two-form and dilaton turned on. 
To examine the solutions, we differentiate
once more and we find
\eqn\derta{
{\partial^2 f_{\alpha} \over \partial \gamma_i \partial \gamma_j }
 =  {1 \over \alpha_+^2}
f_{\beta} (\Gamma_{\lambda^L} \Gamma_{\sigma_j}
\Gamma_{\lambda^L} \Gamma_{\sigma_i} )^{\beta}_{\alpha}
= {1 \over 2 \alpha_+^2 } f_{\beta}
(\Gamma_{\sigma_j}\Gamma_{\sigma_i} )^{\beta}_{\alpha}}
\eqn\dertb{
{\partial^2 f_{\alpha}
\over \partial \gamma_i \partial \tilde{\gamma}_j }
 =  {1 \over \alpha_+ \tilde{\alpha}_+}
f_{\beta} (\Gamma_{\tilde{\lambda}^L} \Gamma_{\tilde{\sigma}_j}
\Gamma_{\lambda^L} \Gamma_{\sigma_i} )^{\beta}_{\alpha} }
\eqn\dertc{
{\partial^2 f_{\alpha}
\over \partial \tilde{\gamma}_i \partial \tilde{\gamma}_j }
 =  {1 \over \tilde{\alpha}_+^2}
f_{\beta} (\Gamma_{\tilde{\lambda}^L} \Gamma_{\tilde{\sigma}_j}
\Gamma_{\tilde{\lambda}^L} \Gamma_{\tilde{\sigma}_i} )^{\beta}_{\alpha}
= {1 \over 2 \tilde{\alpha}_+^2 } f_{\beta}
(\Gamma_{\tilde{\sigma}_j}\Gamma_{\tilde{\sigma}_i} )^{\beta}_{\alpha}}
Because the different $\Gamma$'s anticommute, this leads to the
consistency conditions
\eqn\consi{
f_{\beta}
(\Gamma_{\sigma_j}\Gamma_{\sigma_i} )^{\beta}_{\alpha}
=f_{\beta}
(\Gamma_{\tilde{\sigma}_j}\Gamma_{\tilde{\sigma}_i} )^{\beta}_{\alpha}
=0}

The number of solutions to these equations is counted as follows:
$f_{\beta}$ can either be in the kernel of all $\Gamma_{\sigma_i}$ at
the same time, or it can be in the kernel of all $\Gamma_{\sigma_i}$
except one. using the explicit representation of the $\Gamma$ matrices
above, one sees that there are no other possibilities, and that each
of these possibilities reduces the numbers of supersymmetries by
$2^{-d_1}$. There are $d_1+1$ possibilities though. In a similar
way the tilded sector can be analyzed. The result is that the number
of RR ground states that commute with all screening charges is equal to
\eqn\degen{2^{-d_1} (d_1+1) 2^{-d_2} (d_2+1) 2^{d_1+d_2+1} =
2(d_1+1)(d_2+1) }

 For $AdS_3$ we can make a change of variables
\eqn\chvar{\hat{\sigma}_1=\sigma_1, \quad
\hat{\lambda}_L=\lambda_L-{1\over\alpha_+} \gamma_1 \sigma_1, \quad
\hat{\psi}^L_1=\psi^L_1+ {2\over\alpha_+} \gamma_1 \lambda^L -
{1\over \alpha_+^2} \gamma_1^2 \sigma_1,
\quad \hat{\beta}_1=\beta_1-{2\over\alpha_+} \lambda^L \sigma_1}
This change of variables leaves the OPE's of the fermions invariant,
and it also leaves the OPE's between $\beta$ and the fermions invariant
(i.e. they commute). The screening current
 is now just $e^{-2\phi/\alpha_+}\hat{\beta}$,
and clearly leaves all RR ground states made from hatted fermions
invariant. This is in agreement with the result above where
we found $8$ invariant states for $AdS_3 \times AdS_3$.

It remains to examine whether the vertex operators that create the RR
ground states have the correct OPEs with the $N=1$ supersymmetry generator.
To get no pole in the OPE of $G$ with $V$ of order $3/2$, we need
\eqn\morecon{
{\partial f^{\beta} \over \partial \gamma_i}
(\Gamma_{\psi^L_i})^{\alpha}_{\beta}
+{Q \over 2}f^{\beta} (\Gamma_{\lambda^L})_{\beta}^{\alpha}
+
{\partial f^{\beta} \over \partial \tilde{\gamma}_i}
(\Gamma_{\tilde{\psi}^L_i})^{\alpha}_{\beta}
+{\tilde{Q} \over 2} f^{\beta}(\Gamma_{\tilde{\lambda}^L})_{\beta}^{\alpha} = 0}

Using the previous results, we can rewrite this as

\eqn\moreconn{ {1 \over \alpha_+} (
f^{\beta} (
\Gamma_{\lambda^L} \Gamma_{\sigma_i} \Gamma_{\psi^L_i}
)^{\alpha}_{\beta}
+f^{\beta} (\Gamma_{\lambda^L})_{\beta}^{\alpha} ) =
-{1 \over \tilde{\alpha_+} } ( f^{\beta} (
\Gamma_{\tilde{\lambda}^L} \Gamma_{\tilde{\sigma}_i}
\Gamma_{\tilde{\psi}^L_i})^{\alpha}_{\beta}
+
f^{\beta}(\Gamma_{\tilde{\lambda}^L})_{\beta}^{\alpha} )}

We already know that all $\Gamma_{\sigma_i}$ except at most
one yield zero when acting on $f^{\beta}$, and similarly
for $\Gamma_{\tilde{\sigma}_i}$. Take an $f^{\beta}$
for which all $\Gamma's$ except maybe $\Gamma_{\sigma_1}$
and $\Gamma_{\tilde{\sigma}_1}$ have zero eigenvalue.
We can then rewrite \moreconn\ as
\eqn\morecnn{
f^{\beta} (\Gamma_{\lambda} [\Gamma_{\sigma_1},
\Gamma_{\psi^L_1}] )^{\alpha}_{\beta} =
-{\alpha_+ \over \tilde{\alpha}_+ }
f^{\beta} (\Gamma_{\tilde{\lambda}}
[\Gamma_{\tilde{\sigma}_1},\Gamma_{\tilde{\psi}^L_1}])^{\alpha}_{\beta}
}
After some further manipulations we get
\eqn\fincn{f^{\beta}
([\Gamma_{\lambda},\Gamma_{\tilde{\lambda}}]
 [\Gamma_{\sigma_1},\Gamma_{\psi_1^L}]
 [\Gamma_{\tilde{\sigma}_1},\Gamma_{\tilde{\psi}_1^L}])_{\beta}^{\alpha}
= 8 {\tilde{\alpha}_+ \over \alpha_+ } f^{\alpha} .}
The operator $[\Gamma_{\lambda},\Gamma_{\tilde{\lambda}}]$
has eigenvalues $\pm 2i$, whereas
$[\Gamma_{\sigma_1},\Gamma_{\psi_1^L}]$
and $[\Gamma_{\tilde{\sigma}_1},\Gamma_{\tilde{\psi}_1^L}]$
have eigenvalue $\pm 2$. Thus we see that the eigenvalues are
correlated in some way, depending on whether $\tilde{\alpha}_+ =
+i \alpha_+$ or $-i\alpha_+$. (which was needed in order that
we get the central charge ${c}=3(d_1+d_2+1)$).
This correlation,
which also applies to the the cases when other
$\Gamma_{\sigma_i}$ annihilate $f^{\beta}$, removes
another half of the supersymmetries, leaving us with
$(d_1+1)(d_2+1)$ spacetime supersymmetries.
(Of course, in case $d_1+d_2 < 4$ we still need to add
an internal sector to the theory which may increase
the number of supersymmetries. This only happens for
$d_1+d_2=2$ however, in which case we can get an extra factor
of two.) Finally, the total number of space-time supersymmetries
after combining left- and right movers is twice
the result above, i.e. $2(d_1+1)(d_2+1)$.

The correlation of the signs found above can be naturally embedded
in some GSO projection; there is a subtlety related to the
fact that we have $\gamma$-dependent coefficients. However,
differentiating with respect to some $\gamma$ acts as two gamma
matrices and will not disturb the sign correlation found above.

Let us illustrate this for $AdS_5 \times S^5$. 
This is a theory with ten bosons and fermions and central
charge $c=15$, and can therefore be used as an NSR string
background.  Denote the RR
ground states as usual with five $\pm$ signs.
The first two refer to the eigenvalues of
$[\Gamma_{\sigma_i},\Gamma_{\psi_i^L}]$, the
third one to the eigenvalue of
$[\Gamma_{\lambda},\Gamma_{\tilde{\lambda}}]$
and the last two to the eigenvalues of
$[\Gamma_{\tilde{\sigma}_i},\Gamma_{\tilde{\psi}_i^L}]$.
Then the nine ground states correspond to those combinations
$(\pm,\pm,\pm,\pm,\pm)$ for which the first two signs
are not both minus, the last two signs are not both minus,
and for which the product of all signs is $+1$.
The real RR ground states are, however, still
linear combinations of such states with $\gamma$-dependent
coefficients.

The $AdS_5\times S^5$ theory has a rather rich structure.
 For example, the following five currents are conserved
(i.e. commute with all the screening charges)
\eqn\curlis{\eqalign{
J_1 = & i(\lambda- \frac{1}{\alpha_+} \sigma_r \gamma_r)(
 \tilde{\lambda} - \frac{1}{\tilde{\alpha}_+} 
\tilde{\sigma}_r \tilde{\gamma}_r ) \cr
J_2 = & i(\partial \phi - \frac{1}{\alpha_+}
 \beta_r \gamma_r + \frac{1}{\alpha_+} \sigma_r \psi_r^L) \cr
J_3 = & i(\partial \tilde{\phi} - 
\frac{1}{\tilde{\alpha}_+} \tilde{\beta}_r \tilde{\gamma}_r
+ \frac{1}{\tilde{\alpha}_+} \tilde{\sigma}_r \tilde{\psi}_r^L) \cr
J_4 = & \frac{1}{2} (
\sigma_1 \psi_1^L - \sigma_2 \psi_2^L -
\frac{2}{\alpha_+} \lambda \sigma_1 \gamma_1 +
\frac{2}{\alpha_+} \lambda \sigma_2 \gamma_2 -
\frac{2}{\alpha_+^2} \sigma_1 \sigma_2 \gamma_1 \gamma_2) \cr
J_5 = & \frac{1}{2} (\tilde{\sigma}_1 \tilde{\psi}_1^L
- \tilde{\sigma}_2 \tilde{\psi}_2^L -
 \frac{2}{\tilde{\alpha}_+} \tilde{\lambda} 
\tilde{\sigma}_1 \tilde{\gamma}_1 
 +\frac{2}{\tilde{\alpha}_+} \tilde{\lambda} 
\tilde{\sigma}_2 \tilde{\gamma}_2 
 -\frac{2}{\tilde{\alpha}^2_+}
\tilde{\sigma}_1 \tilde{\sigma}_2 
\tilde{\gamma}_1 \tilde{\gamma}_2 ).}}
In $J_2$ and $J_3$ we recognize a supersymmetric extension of
the current $J_0$ discussed in section~2.3. Out of these five
currents one can construct the following linear combination
\eqn\nistwo{J_{U(1)} = J_1 + 
\frac{2}{\tilde{\alpha}_+} J_2 - \frac{2}{\alpha_+} J_3
+ J_4 + J_5 .}
If we decompose the $N=1$ generator $G$ in terms of its 
$+1$ and $-1$ eigenvalue with respect to $J_{U(1)}$, we find
that $J_{U(1)},G^+,G^-,T$ generate the $c=15$ $N=2$ algebra.
This $N=2$ algebra will probably play a crucial role in the
$AdS_5 \times S^5$ theory, in particular in finding the precise
GSO projection that will give rise to a modular invariant 
string theory. 

\newsec{WZW cosets for supergroups}

It is very interesting to repeat the construction of section 3
for supergroups, especially in view of the results of \bvw\ and
\bzv. Because supergroups contain anticommmuting world-sheet
scalars, at first sight one expects to obtain sigma models 
in the GS formalism. However, we were unable to find $\kappa$
symmetry in the examples below, and furthermore the results of \bvw\ 
indicate that this is probably not the correct interpretation. 

Below we will discuss two examples of super"cosets". Although
the precise embedding of these theories in a critical string
theory remains elusive, both theories have several interesting
features. In particular, both have as target space $AdS_5 \times S^5$
(after analytic continuation).
The first example is the coset $SL(3|3)/SL(2)^2$, the second one
the coset $SL(4|4)/SP(2)^2$. The first one has $18$ anticommuting
scalars, suggesting a close relation with the NSR model of section~4,
the second one has $32$ anticommuting scalars, suggesting a close
relation with the usual GS string. The treatment of $SL(4|4)/SP(2)^2$
will not be quite as satisfactory as that of $SL(3|3)/SL(2)^2$,
because we have to adjust the levels of the two $SL(4)'s$ in 
$SL(4|4)$ at an intermediate stage, but the final result will
still be an exact CFT. 

\subsec{$SL(3|3)/SL(2)^2$}

We basically want to follow the same procedure as we did for
$SL(3)/SL(2)$ in section~3. We again 
write $g = g_ug_{sl(2)^2}g_Dg_l$. To define
the various group elements appearing in this decomposition we
introduce the $3 \times 3$ matrices
\eqn\aone{
A_1 = \pmatrix{ 1 & \bar{\gamma}_1 & \bar{\gamma}_2  \cr
     0 & 1 & 0  \cr
 0 & 0 & 1  \cr}
}
\eqn\atwo{
A_2  =
\pmatrix{1 & 0 & \bar{\gamma}_4  \cr
     0 & 1 & \bar{\gamma}_3\cr
 0 & 0 &  1\cr}
}
\eqn\done{
D_1  = \pmatrix{ e^{2\phi_1} & 0 & 0 \cr
     0 & e^{-\phi_1} & 0 \cr
 0 & 0 & e^{-\phi_1} \cr}
}
\eqn\dtwo{
D_2 = \pmatrix{e^{\phi_2} & 0 & 0 \cr
 0 & e^{\phi_2} & 0\cr
     0 & 0 & e^{-2\phi_2} \cr}
}
\eqn\psione{
\Psi = \pmatrix{
   \bar{\psi}_{11} &  \bar{\psi}_{12} & \bar{\psi}_{13}
\cr
   \bar{\psi}_{21} &  \bar{\psi}_{22} & \bar{\psi}_{23}
\cr
   \bar{\psi}_{31} &  \bar{\psi}_{32} & \bar{\psi}_{33}
\cr}
}

\eqn\decom{
g_u  = \pmatrix{ A_1 & \Psi \cr
 0 & A_2 \cr}
}
\eqn\decomtwo{
g_d = \pmatrix{ D_1 & 0 \cr
 0 & D_2 \cr}
}
\eqn\decomthree{
g_l = \pmatrix{ A_1^{\dagger} & 0 \cr
 \Psi^{\dagger} & A_2^{\dagger} \cr}
}
The two $sl(2)$'s live in the bottom right $2\times 2$ block
of $D_1$, and the top left $2 \times 2$ block of $D_2$. After
removing the $sl(2)^2$ degrees of freedom, the remaining
action is given by the WZW action evaluated on $g=g_u g_D g_l$.
In addition, we need a linear dilaton background in order
to make the theory exactly conformal. We will determine this
dilaton background later. The WZW action evaluated for
$g=g_u g_D g_l$ reads
\eqn\actionf{\eqalign{
S  &= {3 k \over 2 \pi}  \int (\partial \phi_1 \bar{\partial} \phi_1
 - \partial \phi_2 \bar{\partial} \phi_2  ) + \cr
&  {k \over 2\pi} \int {\rm tr}((
\partial A_1 D_1 \bar{\partial} A_1^{\dagger} D_1^{-1})  -
(\partial A_2 D_2 \bar{\partial} A_2^{\dagger} D_2^{-1} ))  + \cr
& {k \over 2\pi} \int {\rm tr} (
(A_1 D_1 A_1^{\dagger})^{-1} \partial (\Psi A_2^{-1}) (A_2 D_2
A_2^{\dagger})
\bar{\partial} ( (\Psi A_2^{-1})^{\dagger} ))} .
}
Upon expanding the first two lines in $\phi_i,\chi_i,\bar{\chi}_i$
we see that they describe two copies of the $AdS_5$ sigma
model discussed in section~2. The last line in  \actionf\ is a
term bilinear in the anticommuting scalars $\Psi,{\Psi}^{\dagger}$.
The signs arise because the action for $SL(3|3)$
contains a supertrace rather than an
ordinary trace. Furthermore we notice that the first two
lines of \actionf\ 
can be combined to form $kW(A_1D_1A_1^{\dagger}) -
kW(A_2 D_2 A_2^{\dagger})$, with $W$ the WZW action for $SL(4)$.

The next step in the construction is to bosonize \actionf.
We will do this in two steps. The first step involves the
anticommuting world-sheet scalars $\Psi,\Psi^{\dagger}$,
the second step involves only the bosons of the theory.

To perform the first step we start by redefining the field $\Psi$ 
as $\Psi A_2^{-1}=\tilde \Theta$
in order to simplify the
last term in \actionf.
Next we redefine
$\tilde \Theta = M_1 \Theta M_2^{-1}$ with $M_1 = A_1 D_1 A_1^+,
M_2 = A_2 D_2 A_2^+$.  This changes the last term in \actionf\
into:
\eqn\redef{(\partial \Theta +M_1^{-1}\partial M_1 \Theta - \Theta
M_2^{-1}\partial M_2)\bar \partial \Theta^{+}}
This can be transformed in the action for a fermionic
$\beta,\gamma$-system by introducing the new field
\eqn\newf{\beta_{\Theta} = \partial \Theta +M_1^{-1}\partial M_1 \Theta -
\Theta M_2^{-1}\partial M_2 .}

 From this expression we see that $\Theta$ transforms under the action of
$SL(3) \times SL(3)$ into $M_1^{-1}\Theta M_2$ and therefore the
Jacobian
for the change of variables 
\newf\ is given by the WZW actions for $M_1$ and $M_2$,
computed in the bi-fundamental representation. Since the bi-fundamental
representation of $sl(3) \times sl(3)$ contains 3 fundamental
representations
of each $sl(3)$, we obtain the following correction to the action

\eqn\integr{ + 3 W((A_1 D_1 A_1^{\dagger})) +  3 W(A_2 D_2 A_2^{\dagger})}

This, combined with the first two terms in our action, leads
to a renormalisation of the coupling in front of these terms.
The full action at this stage reads
\eqn\integrr{ (k+3) W((A_1 D_1 A_1^{\dagger})) +
(-k+3) W(A_2 D_2 A_2^{\dagger}) + {k \over 2\pi} \int \beta_{\Theta}
\bar{\partial} \Theta^{\dagger}        }

Having bosonized the anti-commuting degrees of freedom, it remains
to bosonize the bosonic degrees of freedom. This can be done
exactly as explained in section~3. The first two terms in
\integrr\ correspond to two $SL(3)/SL(2)$
versions of our ``coset'' construction, one with level $k+3$
and another with level $-k+3$. Thus we can use the results in
section~3.2
and conclude that the shift by $3$ gets canceled.
We also obtain
two bosonic screening charges for each $SL(3)$ of the form given in
\scr. After a rescaling of the fields the system is described by
the action
\eqn\freeact{\eqalign{S = 
{1\over 4\pi}
\int ((\partial \phi_1 \bar{\partial} \phi_1
 - \partial \phi_2 \bar{\partial} \phi_2  ) + Q_1 R\phi_1 + Q_2 R\phi_2
\cr
+ \beta_i^1 \bar \partial \gamma_i^1 + \beta_1^2 \bar \partial \gamma_i^2+
\beta_{\Theta} \bar \partial \Theta^+ ) } }
with
\eqn\chrg{Q_1=\sqrt{3 \over 8k}, \qquad Q_2 = \sqrt{3 \over 8k} }

The change of variables for both the anti-commuting and the
commuting degrees of freedom generate screening charges.
The bosonic ones are obtained from \scr, and in the normalization
\freeact\ they are given by
\eqn\screenings{
 S_i=\int \beta_i e^{\sqrt{3/2k}\phi_1}, \quad i=1,2,\quad
 S_i=\int \beta_i e^{\sqrt{3/2k}\phi_2}, \quad i=3,4 .}

These screening charges do not have conformal weight equal
to zero, because the charges $Q_i$ are equal to twice the
momenta in \screenings . Therefore this theory is not
conformal. This should not come as a surprise, because we already
knew from the $SL(3)/SL(2)$ example in section~2 that the
only way to make sense out of the theory is add a suitable linear
dilaton background from the start.  Here we see that we should have
added the linear dilaton background
\eqn\lindil{\Phi = -{Q_1 \over 2}\phi_1 - {Q_2 \over 2} \phi_2}
at the beginning to make the theory conformal. The final
exact CFT is therefore given by \freeact\ with $Q_i$ replaced by $Q_i/2$,
it has central charge
\eqn\mocen{c = 5 + 12 {3 \over 32k} + 5 - 12 {3 \over 32k}-18=-8}
and it has the screening charges \screenings, plus certain 
fermionic screening charges.
The latter originate from the change of variables \newf:
\eqn\scrrr{M_1 \beta_{\Theta} M_2^{-1} =\partial (M_1 \Theta M_2^{-1})}

Some of the fermionic screening charges are simple, and some of
them are iterated. Following the discussion in section~3.2,
we will only present the simple fermionic screening 
charges and verify
whether they have conformal weight one. 
This is a nontrivial statement, as we already used all
available freedom to change the background charges of the
scalars in order to give the bosonic screening charges
the right conformal weight.

The fermionic screening charges, as is clear from \scrrr , are the
contour integrals of $M_1 \beta_{\Theta} M_2^{-1}$.
Let us denote the components of $\beta_{\Theta}$ by $\beta_{ij}$,
and the $i,j$ component of $M_1 \beta_{\Theta} M_2^{-1}$ by
$S_{ij}$. Then out of a total of nine fermionic screening 
currents we 
find the following four simple ones
charges
\eqn\fermsc{\eqalign{
S_{21} = & e^{(-\phi_1-\phi_2)/\sqrt{6k}}
(\beta_{21} + \beta_{11} \gamma_1
 -\beta_{23} \gamma_4 - \beta_{13} \gamma_1 \gamma_4) \cr
S_{22} = & e^{(-\phi_1-\phi_2)/\sqrt{6k}}
(\beta_{22} + \beta_{12} \gamma_1 -
\beta_{23} \gamma_3 - \beta_{13} \gamma_1 \gamma_3) \cr
S_{31} = & e^{(-\phi_1-\phi_2)/\sqrt{6k}}
(\beta_{31} + \beta_{11} \gamma_2
 -\beta_{33} \gamma_4 - \beta_{13} \gamma_2 \gamma_4) \cr
S_{32} = & e^{(-\phi_1-\phi_2)/\sqrt{6k}}
(\beta_{32} + \beta_{12} \gamma_2 - \beta_{33} \gamma_3 -
\beta_{13} \gamma_2 \gamma_3) }}

The conformal weight of $\exp(p_1 \phi_1 + p_2 \phi_2)$
equals $-p_1(p_1+Q_1)/2 + p_2(p_2+Q_2)/2$, and since
$Q_1=Q_2$ this vanishes when $p_1=p_2$. Therefore all
four fermionic screening charges have the right conformal weight..

This completes the construction of the CFT announced in the introduction
for the case of $SL(3|3)/SL(2)^2$. It has $AdS_5 \times S^5$ target
space, 18 anticommuting world-sheet scalars 
and bosonic and fermionic $B$-fields turned
on. We have shown that theory is exactly conformal and can be bosonised
with the help of the free fields and screening operators listed above.

\subsec{A Comment on the Bosonization of $SL(N|N)$}

The change of variables \newf\ can also be applied
to the $SL(N|N)$ WZW theory. The result is an action
of the form
\eqn\interm{S=(k+N)W(g_1) + (-k+N)W(g_2) + \frac{k}{2\pi}
 \int d^2 z \beta_{\Theta} \bar{\partial} \Theta^{\dagger} }
where $g_1$ and $g_2$ are arbitrary $sl_3$ group elements,
and $\beta_{\Theta},\Theta$ are anticommuting $N\times N$
matrices. Explicit expressions for the $SL(N|N)$ currents
in terms of the variables in \interm\ are given in
\bars.  The central charge of \interm\ is
\eqn\checkcent{ c= { (k+N) (N^2-1) \over k+N-N } +
 { (-k+N) (N^2-1) \over -k+N-N } - 2 N^2 = -2}
which is indeed the correct result for $SL(N|N)$. 
By adding two more free scalar fields we find a theory
that has $c=0$. this is the $GL(N|N)$ theory that
was studied as a topological field theory in 
\ref\isra{J.M. Isidro and A.V. Ramallo, ``$gl(n,n)$
Current Algebras and Topological Field Theories,''
Nucl. Phys. {\bf B414} (1994) 715, hep-th/9307037.}.
The change of variables above maps it to the topological
$GL(N)/GL(N)$ theory, although this is not an exact
equivalence, because the change of variables introduces
certain fermionic screening charges in the theory. 
Along the same lines, the $SL(N|N)$ WZW theory is
can be mapped to the topological $SL(N)/SL(N)$ theory
together with one additional free anticommuting $\beta,\Theta$
system.

In the case of $SL(2|2)$, the map above is essentially
the one discussed in section~10 of \bvw, namely it
maps the $SL(2|2)$ WZW model to two copies of the $SL(2)$
WZW model plus four anticommuting $\beta_{\Theta},\Theta^{\dagger}$
systems. A further change of variables then maps it to
the standard NSR description of strings on $AdS_3\times
S^3$ with only NS background fields turned on.

\subsec{$SL(4,4)/SP(2)^2$}

Because the GS string on $AdS_5 \times S^5$ is described by
a sigma model whose target space is (a suitable real form of)
the coset superspace $SL(4,4)/SP(2)^2$, it is very interesting
to examine what happens when we examine our ``coset'' for
$SL(4,4)/SP(2)^2$. One novelty here is that we don't know
whether it is possible to express $SL(4)$ currents in terms
of $SP(2)$ currents and additional free scalars and $\beta,\gamma$
systems. This is because $SP(2)$ is not generated by a subset
of the positive and negative simple roots, and therefore the
results in \df\ cannot be used. Still, we will try to construct
an exact CFT by repeating the procedure we have been following
so far. We write $g=g_ug_{sp(2)^2}g_dg_l$, and then remove
the $SP(2)^2$ degrees of freedom. It will turn out that in 
order to obtain an exact CFT, we not only need to adjust the
background charges, but we also need to adjust the levels
of some of the WZW actions that appear in the discussion. This
is probably related to the absence of a realization of $SL(4)$
currents in terms of $SP(2)$, as mentioned above.

Except for this, the construction of the exact CFT follows
closely the construction for $SL(3|3)/SL(2)^2$. We begin by
defining $g_u,g_d,g_l$ in terms of the auxiliary 
$4 \times 4$ matrices
\eqn\paonee{
A_1 = \pmatrix{ 1 & \bar{\gamma}_1 & \bar{\gamma}_2 & 0 \cr
     0 & 1 & 0 & 0 \cr
 0 & 0 & 1 & 0 \cr
     0 & 0 & 0 & 1  \cr}
}
\eqn\patwoe{
A_2  =
\pmatrix{1 & 0 & 0 & 0 \cr
     0 & 1 & 0 & \bar{\gamma}_4 \cr
 0 & 0 & 1 & \bar{\gamma}_3 \cr
     0 & 0 & 0 & 1  \cr}
}
\eqn\pdonee{
D_1  = \pmatrix{ e^{3\phi_1} & 0 & 0 & 0 \cr
     0 & e^{-\phi_1} & 0 & 0 \cr
 0 & 0 & e^{-\phi_1} & 0 \cr
     0 & 0 & 0 & e^{-\phi_1}  \cr}
}
\eqn\pdtwoe{
D_2 = \pmatrix{ e^{\phi_2} & 0 & 0 & 0 \cr
     0 & e^{\phi_2} & 0 & 0 \cr
 0 & 0 & e^{\phi_2} & 0 \cr
     0 & 0 & 0 & e^{-3\phi_2} \cr}
}
\eqn\ppsionee{
\Psi = \pmatrix{
   \bar{\psi}_{11} &  \bar{\psi}_{12} & \bar{\psi}_{13} &  \bar{\psi}_{14}
\cr
   \bar{\psi}_{21} &  \bar{\psi}_{22} & \bar{\psi}_{23} &  \bar{\psi}_{24}
\cr
   \bar{\psi}_{31} &  \bar{\psi}_{32} & \bar{\psi}_{13} &  \bar{\psi}_{34}
\cr
   \bar{\psi}_{41} &  \bar{\psi}_{42} & \bar{\psi}_{43} &  \bar{\psi}_{44}
\cr}
}
The WZW action $kW(g_ug_dg_l)$ is evaluated by taking
\eqn\decome{
g_u  = \pmatrix{ A_1 & \Psi \cr
 0 & A_2 \cr}
}
\eqn\decomtwoe{
g_d = \pmatrix{ D_1 & 0 \cr
 0 & D_2 \cr}
}
\eqn\decomthreee{
g_l = \pmatrix{ A_1^{\dagger} & 0 \cr
 \Psi^{\dagger} & A_2^{\dagger} \cr}
}
and this leads to an action which is very similar to \actionf :
\eqn\actionfe{\eqalign{
S  &= {3 k \over \pi}  \int (\partial \phi_1 \bar{\partial} \phi_1
 - \partial \phi_2 \bar{\partial} \phi_2  ) + \cr
&  {k \over 2\pi} \int {\rm tr}((
\partial A_1 D_1 \bar{\partial} A_1^{\dagger} D_1^{-1})  -
(\partial A_2 D_2 \bar{\partial} A_2^{\dagger} D_2^{-1} ))  + \cr
& {k \over 2\pi} \int {\rm tr} (
(A_1 D_1 A_1^{\dagger})^{-1} \partial (\Psi A_2^{-1}) (A_2 D_2
A_2^{\dagger})
\bar{\partial} ( (\Psi A_2^{-1})^{\dagger} ))} .
}
Again, the first two lines are just $kW(A_1D_1A_1^{\dagger}) -
kW(A_2 D_2 A_2^{\dagger})$, with $W$ the WZW action.
However, in contrast to the case of 
$SL(3|3)/SL(2)^2$, it turns out that the classical action \actionf\
does not quite correspond to an exact CFT. We have to slightly 
modify it by hand, namely in order to obtain an exact CFT 
we should adjust the levels in \actionfe\ to
\eqn\actionffe{\eqalign{
S  &=
(k-{4\over 3} ) W(A_1D_1A_1^{\dagger}) +
(-k-{4\over 3}) W(A_2 D_2 A_2^{\dagger}) + \cr
& {k \over 2\pi} \int {\rm tr} (
(A_1 D_1 A_1^{\dagger})^{-1} \partial (\Psi A_2^{-1}) (A_2 D_2
A_2^{\dagger})
\bar{\partial} ( (\Psi A_2^{-1})^{\dagger} ))}
}
These shifts of $k$ look rather awkward. Perhaps there is a better
way to obtain the same final answer without making such shifts in the levels.
At this point we lack a better geometrical description of this ``coset''.
We will comment on this later. 

To continue the bosonization procedure, we redefine the field
$\Psi$ exactly as in \redef\ and \newf . In this case,
$\Theta$ transforms in the bifundamental representation
of $SL(4) \times SL(4)$ which contains four fundamentals
for each $SL(4)$. This leads to a correction to the
action of the form
\eqn\integre{ + 4 W(A_1 D_1 A_1^{\dagger}) +  4 W(A_2 D_2 A_2^{\dagger})}
and the  full action at this stage reads
\eqn\integrre{ (k+{8 \over 3}) W(A_1 D_1 A_1^{\dagger}) +
(-k+{8 \over 3}) W(A_2 D_2 A_2^{\dagger}) + {k \over 2\pi} \int \beta_{\Theta}
\bar{\partial} \Theta^{\dagger}        }
 
Next, we bosonize the bosonic part of the action. This is given by
the $SL(4)/SP(2)$ version of our general construction. This
is similar to $SL(4)/SL(3)$, except that there are 
only two rather than three $\gamma,\bar{\gamma}$ pairs.
Correspondingly, the shift
of $k$ will not be by $-4$
but only by $-8/3$\foot{The fact that the shift here is $-8/3$ as opposed to
$-4$ explains the adjustment we had to make in order to go from
\actionfe\ to \actionffe . If we would be able to express $SL(4)$
currents in terms of $SP(2)$ currents and free $\phi,\beta,\gamma$ fields,
there probably would be no need to adjust levels by hand and the
whole construction would appear more natural. However, at present
we do not know either algebraically or in the Lagrangian approach
whether this is possible.}.
As a result, the various $8/3$'s cancel, and after a suitable
rescaling of the fields we obtain the free field action
\eqn\freeacte{\eqalign{S = {1\over 4\pi} \int 
((\partial \phi_1 \bar{\partial} \phi_1
 - \partial \phi_2 \bar{\partial} \phi_2  ) + Q_1 R\phi_1 + Q_2 R\phi_2
\cr
+ \beta_i^1 \bar \partial \gamma_i^1 + \beta_1^2 \bar \partial \gamma_i^2+
\beta_{\Theta} \bar \partial \Theta^+)}}
with background charges
\eqn\chrge{Q_1={4 \over\sqrt{3k}}, \qquad Q_2 = {4 \over \sqrt{3k}} }

Now everything is practically the same as before. The various changes of
variables give rise to bosonic and fermionic screening charges.
The explicit expressions for the simple screening charges
 will be given in
a moment. In order that the screening charges have conformal
weight zero, we need to replace $Q_i$  by $Q_i/2$ in \freeacte .
Thus the final exact CFT is given by \freeacte , with 
$Q_i=2/\sqrt{3k}$. It has central charge $-22$, four bosonic screening
charges and sixteen fermionic screening charges, of which nine are
simple and seven are iterated. The bosonic screening charges are
\eqn\screeningse{
 S_i=\int \beta_i e^{2 \phi_1/\sqrt{3k}}, \quad i=1,2,\quad
 S_i=\int \beta_i e^{2 \phi_2/\sqrt{3k}}, \quad i=3,4 .}

To write down the simple fermionic screening charges, we 
denote the components of $\beta_{\Theta}$ by $\beta_{ij}$,
and the $i,j$ component of $M_1 \beta_{\Theta} M_2^{-1}$ by
$S_{ij}$. The nine simple fermionic screening
currents are given by
\eqn\fermsce{\eqalign{
S_{21} = & e^{(-\phi_1-\phi_2)/\sqrt{12k}}
(\beta_{21} + \beta_{11} \gamma_1) \cr
S_{22} = & e^{(-\phi_1-\phi_2)/\sqrt{12k}}
(\beta_{22} + \beta_{12} \gamma_1 -
\beta_{24} \gamma_4 - \beta_{14} \gamma_1 \gamma_4) \cr
S_{23} = & e^{(-\phi_1-\phi_2)/\sqrt{12k}}
(\beta_{23} + \beta_{13} \gamma_1 - \beta_{24} \gamma_3
-\beta_{14} \gamma_1 \gamma_3) \cr
S_{31} = & e^{(-\phi_1-\phi_2)/\sqrt{12k}}
(\beta_{31} + \beta_{11} \gamma_2) \cr
S_{32} = & e^{(-\phi_1-\phi_2)/\sqrt{12k}}
(\beta_{32} + \beta_{12} \gamma_2 - \beta_{34} \gamma_4 -
\beta_{14} \gamma_2 \gamma_4) \cr
S_{33} = & e^{(-\phi_1-\phi_2)/\sqrt{12k}}
(\beta_{33} + \beta_{13} \gamma_2 - \beta_{34} \gamma_3
-\beta_{14} \gamma_2 \gamma_3) \cr
S_{41} = & e^{(-\phi_1-\phi_2)/\sqrt{12k}}
\beta_{41} \cr
S_{42} = & e^{(-\phi_1-\phi_2)/\sqrt{12k}}
(\beta_{42} - \beta_{44} \gamma_4 ) \cr
S_{43} = & e^{(-\phi_1-\phi_2)/\sqrt{12k}}
(\beta_{43} - \beta_{44} \gamma_3 ) . }}

This concludes the description of the exact CFT based on $SL(4|4)/SP(2)^2$.

\newsec{Conclusions}

In this paper we considered various conformal field theories whose
target space geometry is an anti-de Sitter space. We have not yet
achieved a full understanding of these CFT's. One issue that 
requires a precise justification is that of simple versus iterated
screening charges. We assumed by analogy with WZW theory (when the rank of
the group is greater than one) that the
conformal blocks of the NSR version of the $AdS_{2d+1}$ sigma model
can be computed using only simple screening charges. A deeper
understanding of this assumption and the way the left and right
movers should be combined is certainly desirable, in particular
in order to construct a complete modular invariant theory with
a proper GSO projection. 
In this context we would like to note the
similarities with $c<1$ theories (Minimal Models) 
coupled to $c>1$ Liouville theory, where the non-compact part of our target
space plays the role of the
Liouville model and the compact part (i.e. the part which is
analytically continued in $\phi$) corresponds to the Minimal Model. 
However,
the quantization of strings on $AdS$ spaces will undoubtedly lead
to the same problems that one encounters when one studies string
theory on $SL(2,R)$ 
\ref\sltworef{
J. Balog, L. O'Raifeartaigh, P. Forgacs and A. Wipf, ``Consistency of
String Propagation on Curved Space-Times: An SU(1,1) Based Counterexample,''
Nucl. Phys. {\bf B325} (1989) 225;
P.M.S. Petropoulos, ``Comments on SU(1,1) String Theory,'' Phys. Lett. 
{\bf B236}
(1990) 151;
N. Mohameddi, ``On the Unitarity of String Propagation on
$SU(1,1)$,'' Int. J. Mod. Phys. {\bf A5} (1990) 3201;
I. Bars and D. Nemeschansky, ``String Propagation in
Backgrounds with Curved Space-Time,'' Nucl. Phys. {\bf B348} (1991) 89;
S. Hwang, ``No-Ghost Theorem for SU(1,1) String Theories,'' Nucl. Phys. 
{\bf B354}
(1991) 100;
M. Henningson and S. Hwang, ``The Unitarity of $SU(1,1)$ Fermionic
Strings,'' Phys. Lett. {\bf B258} (1991) 341;
M. Henningson, S. Hwang, P. Roberts, and B. Sundborg,
 ``Modular Invariance of
$SU(1,1)$ Strings,'' Phys. Lett. {\bf B267} (1991) 350;
S. Hwang, ``Cosets as Gauge Slices in $SU(1,1)$ Strings,''
Phys. Lett. {\bf B276} (1992) 451, hep-th/9110039;
I. Bars, ``Ghost-Free Spectrum of a Quantum String in $SL(2,R)$ Curved
Space-Time,'' Phys. Rev. {\bf D53} (1996) 3308, hep-th/9503205;
``Solution of the $SL(2,R)$ String in Curved Spacetime,''
in ``Future Perspectives In String Theory,'' Los Angeles, 1995,
hep-th/9511187;
Y. Satoh, ``Ghost-free and Modular 
Invariant Spectra of a String in SL(2,R) and Three 
Dimensional Black Hole Geometry,'' Nucl. Phys. {\bf B513} 
(1998) 213, hep-th/9705208;
J. M. Evans, M. R. Gaberdiel, and M. J. Perry, ``The No-Ghost Theorem
for ${\rm AdS}_3$ and the Stringy Exclusion Principle,'' hep-th/9806024;
``The No-Ghost Theorem and Strings on ${\rm AdS}_3$,'' hep-th/9812252.}.

Although the coset theories for supergroups discussed in section~6
look very suggestive, it is not yet clear how to use them to construct
consistent string backgrounds. Based on the number of supersymmetries,
the $SL(3|3)/SL(2)^2$ model appears related to the NSR string on
$AdS_5 \times S^5$ of section~4, whereas the $SL(4|4)/SP(2)^2$
appears related to the ``usual'' string theory on $AdS_5 \times S^5$
with RR five-form flux. It would be very interesting to understand
these relations in more detail. In particular, one could look for
additional world-sheet symmetries, or for an analogue of 
$\kappa$-symmetry.
Once the space-time interpretation
of these theories is understood, one could analyze whether they
do in fact describe theories with some RR background turned on
(as we already mentioned, 
we suspect that RR backgrounds are related to fermionic screening charges)
or whether one needs to perturb the theories away from the 
``WZW'' point in order to turn on RR backgrounds. 

The theories we have described in this paper do have holographic 
behavior, and this raises the question which boundary quantum
field theories are associated to them. To answer this question,
it would be helpful to know to which brane configurations 
they correspond. The $AdS_{2d_1+1} \times S^{2d_2+1}$ solutions
appear to be related to peculiar delocalized superpositions
of intersecting fundamental strings and NS fivebranes. 
The meaning of these configurations is, however, rather obscure.

To summarize, we have described various exact CFT's with $AdS$ target
spaces, some of which can be used to build critical string backgrounds.
We believe that a further study of these theories will teach us more 
about exactly solvable CFT's and their space-time interpretation,
as well as the world-sheet formulation of AdS/CFT duality.

\newsec{\bf Acknowledgements} We would like to thank the organizers of
the Amsterdam Workshop on M-theory and Black Holes in July 1998 where this
collaboration was
initiated. JdB would like to thank the CERN Theory Division and the Aspen Center
for Physics and S.~S. would like to thank the CERN Theory Division, IHES
and the String Theory group at Harvard University, 
for hospitality. In addition, we would like to thank
M.~Bershadsky for collaboration at the initial stage of this project,
and J. Bernstein, E. Frenkel,
N. Nekrasov, H. Ooguri,
Y. Oz, A. Polyakov, C. Sfetsos, C. Vafa, E. Verlinde and E. Witten
for discussions.
The research of S.~S. is supported by DOE grant
DE-FG02-92ER40704, by NSF CAREER award, by OJI award from DOE and by
Alfred P.~Sloan foundation.

\listrefs
\bye